\PassOptionsToPackage{hyphens}{url}

\documentclass[
	12pt,
	superscriptaddress,
	showpacs,
	showkeys,
	twocolumn,
	prb,
	aps,
	reprint,
	a4paper,
	floatfix,
	]{revtex4-1}

\usepackage{graphicx}
\usepackage{amssymb,amsmath}
\usepackage{subcaption}
\usepackage{bm}
\usepackage{tikz}
\usepackage{xstring}
\usepackage[colorlinks=true, allcolors=blue]{hyperref}
\usepackage[mode=math,range-units=single]{siunitx}

\usepackage{upgreek}

\newcommand*\crit{\mathop{}\!\mathrm{c}}
\newcommand*\Sl{\mathop{}\!\mathrm{S}}
\newcommand*\Sp{\mathop{}\!\mathrm{S'}}
\newcommand*\Spp{\mathop{}\!\mathrm{S''}}
\newcommand*\nn{\mathop{}\!\mathrm{n}}

\newcommand*\DIb{\mathop{}\!\Delta I_{\mathrm{bias}}}
\usepackage{natbib}
\begin{document}
\title[Magnetic Field Sensitivity of TESs]{Magnetic Field Sensitivity of Transition Edge Sensors}

\author{R. C. Harwin}
\email{rch66@cam.ac.uk}
\author{D. J. Goldie}
\author{C. N. Thomas}
\author{S. Withington}
\affiliation{Cavendish Laboratory, JJ Thomson Avenue, Cambridge CB3 OHE, United Kingdom.}
\date{\today}

\begin{abstract}
Understanding the magnetic field sensitivity of Transition Edge Sensors (TESs) is vital in optimising the configuration of any magnetic shielding as well as the design of the TESs themselves.
An experimental system has been developed to enable the investigation of the applied magnetic field direction on TES behaviour, and the first results from this system are presented.
In addition, measurements of the effect of applied magnetic field magnitude on both supercurrent and bias current are presented.
The extent to which the current theoretical framework can explain the results is assessed and finally, the impact of this work on the design of TESs and the design of magnetic shielding is discussed.
\end{abstract}


\maketitle

\section{Introduction}\label{sec:intro}

Transition edge sensors (TESs) have become a crucial technology in a number of fields of research.
TESs are used as high-resolution energy-resolving detectors 
for X-rays\,\cite{Gottardi2016,Smith_2021}, optical photons\,\cite{Smith2012a}, electrons\,\cite{Patel2021}, neutrinos\,\cite{Puiu2018} and dark matter candidates\,\cite{Roth2009} in areas such as particle physics,  astronomy and quantum cryptography.
They are also used as highly sensitive power detectors for photon fluxes in mm, sub-mm and far-infrared astronomy\,\cite{Goldie2016,Farias_2022}. 
As a result, there is significant interest in understanding factors that can affect and degrade TES operation.

One such factor is magnetic field sensitivity. 
Magnetic  field sensitivity means TESs must be shielded from stray and/or time-varying fields that might arise in operation.
Sources include electric currents, both in the surrounding wiring and in the TESs themselves \cite{Swetz2012}, components of the cooling system such as cryo-compressors \cite{Bergen2016} or adiabatic demagnetisation refrigerators (ADRs) \cite{Jackson2013}, or other parts of the instrument, for example the beam deflectors in a telescope or motors driving rotating polarizing plates.

 Trade-offs are involved in the design of the shielding.
For example, to couple the TES detector to the desired signal (photons, x-rays etc.), to allow electrical connection to  the external warmer electronics,
or to allow thermal connection to the cooling platform, the magnetic shielding must be engineered to include apertures, through which static and low frequency magnetic fields may penetrate\cite{Jackson2015}. 
In space-missions the level of shielding may have to be traded-off against weight.
In these situations, it is critical to understand the magnetic field sensitivity of TESs in detail, 
so that performance of the design can be assessed and optimized.
The scale of the field sensitivity is obviously critical to determine minimum shielding levels.
Directionality of the sensitivity is also important, e.g. in constraining the orientation of apertures for optical access or  wiring or the routing of harnesses with respect  to the TES. In other circumstances it may be desirable to operate TESs in a magnetic field when an understanding of the likely effect of the magnitude and orientation of the field is also crucial. 
By contrast, recent measurements show that TESs are insensitive to applied electric fields\cite{Patel2023e_field}.
Here we report a systematic experimental study of the effect of applied magnetic field and its orientation on a range on TES operating parameters and derived characteristics (critical current, bias current, transition temperature, power-to-current responsivity) as a function of TES lateral dimensions and geometry. 

A TES is sensitive to magnetic field by virtue of its geometry. At a fundamental level, a TES  comprises a superconductor $\Sp$, often chosen to have with a superconducting critical temperature $T_{\crit} < 1\,\text{K}$,  connected to readout electronics by thinfilm wiring made from a superconductor $\Sl$ having a higher $T_{\text {c}}$.
The TES  operates within the superconducting-normal resistive transition of $\Sp$.
$\Sp$ is typically a superconducting thin-film or superconductor-normal metal (S/N) bilayer with thickness $\sim$ 100\,nm, and lateral dimensions 
$\sim$ 10\,$\upmu$m.
Wiring layer $\Sl$ is a higher  $T_{\text{c}}$ superconductor typically Nb such that $T_{\text{c, wire}}\sim 8{\text K}$.
The combination $\Sl\Sp\Sl$ forms a superconducting weak-link due to the superconducting proximity effect\,\cite{Kozorezov2011}.
In many practical TES designs additional normal metal structures are added on top of $\Sp$ in order to improve other aspects of behaviour such as structures along the longitudinal edges (`banks') to address possible imperfections or 
lateral `bars' added across the direction of current flow.
Previous work has shown that adding banks can reduce B-field sensitivity\,\cite{Sadleir2011}, as does reducing the spacing of any bars\,\cite{Smith2014}.

Observation of the Josephson effect above the intrinsic transition temperature of $\Sp$ in TESs is now well-established 
experimentally\,\cite{Sadleir2010,Sadleir2011}.
It is also well-established that both the effective $T_\text{c}$ and superconducting critical current $I_{\crit}$ in weak links are sensitive to the magnitude of a magnetic field (B) applied in the direction  normal to the plane of the  TES.
Measurements of $I_{\crit}$ with $B$   perpendicular to  the TES exhibit an oscillatory, Fraunhofer-like dependence.
Broadly similar oscillations have been observed in the bias current of the TES $I_{b}$ under voltage-bias\cite{Smith2013,Chiodi2012,Hijmering2014}.
The small-signal electrothermal parameters that characterise the change in TES resistance ($R$) with temperature ($T$) and current ($I$), $\alpha= (T/R) \,\partial R/\partial T$ and $\beta = (I/R) \, \partial R/\partial I $ \cite{Irwin2005} also exhibit oscillatory behaviour with magnetic field \cite{Harwin2017}.
Changes in  $\alpha$ and $\beta$ are expected to  change the small-signal power-to-current responsivity, and therefore the noise equivalent power for bolometric applications and the achievable energy resolution for calorimetric applications \cite{Smith2013}.
A review of the current understanding of TES physics is given by Gottardi et al. \cite{Gottardi2021}.

This paper presents an experimental study of the effect of magnetic field magnitude and orientation on a set of MoAu TESs as a function of TES lateral dimensions.
Section \ref{Sec:Experimental} describes the experimental method, including a description of the TES geometry, the 3-axis magnetic field system, device materials and fabrication processes. 
Section \ref{Sec:Results} gives details of the measurements. For convenience, these are categorised by the type of measurement carried out.
Section \ref{Sec:Effect_of_field_on_Jc} shows the effect on supercurrent of a magnetic field 
applied perpendicular to the plane of the TES.
Section \ref{Sec:Effect_of_temperature_on_Jc} then presents the temperature dependence of the supercurrent for a subset of the TESs studied here, showing the effects of device size.
Section \ref{Sec:Effect_of_field_on_voltage_biased TESs} describes measurements of the effect of applied field on the  current in the TESs studied here under voltage bias, including a first-order estimate of the effect on TES responsivity.
Section \ref{Sec:Directional_dependence} gives details of measurements of the dependence of the bias current of  voltage-biased TESs as a function of the orientation of an applied magnetic field.
Section \ref{Sec:Discussion} discusses the measurements and concludes with recommendations for the use of TESs in a magnetic field. 

\section{Experimental method}\label{Sec:Experimental}
 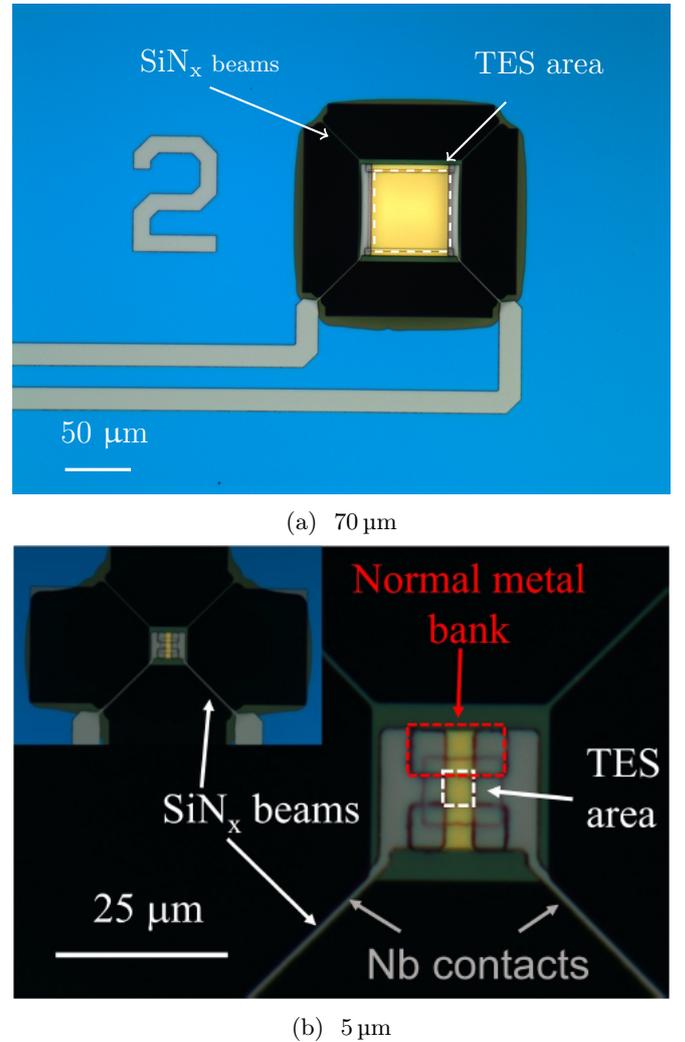
\begin{figure}
\begin{subfigure}{\columnwidth}
\begin{tikzpicture}
\begin{scope}
    \node[anchor=south west,inner sep=0] (image) at (0,0) {\includegraphics[width=\textwidth,angle=0]{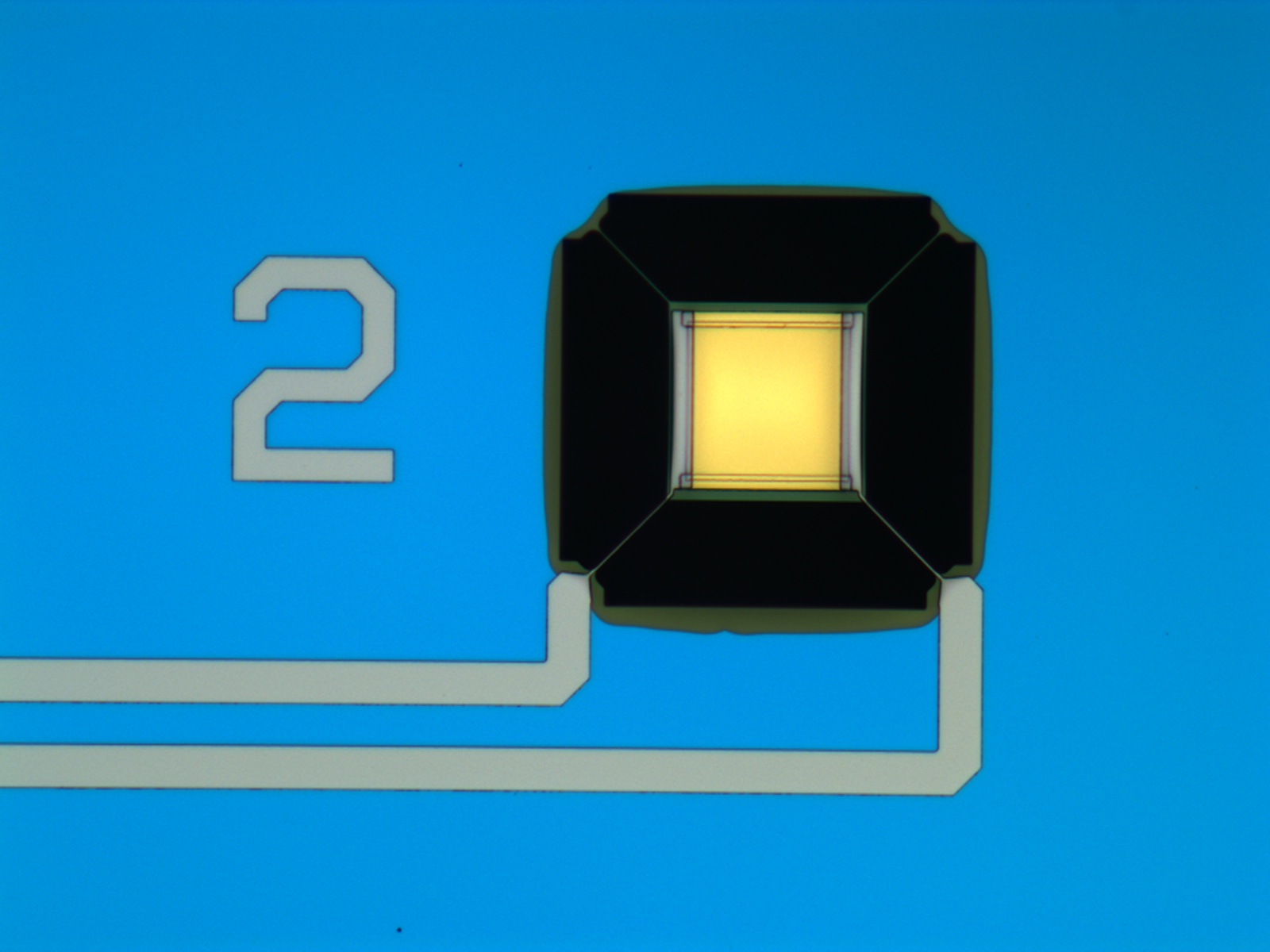}};
    \begin{scope}[x={(image.south east)},y={(image.north west)}]
        \draw[very thick, white, -] (0.08,0.05) -- (0.18,0.05);
        \node at (0.14,0.12) {\color{white}{\large 50 {$\mathrm\upmu$m}}};
        \draw[thick, white, dashed] (0.55,0.66) -- (0.665,0.66) -- (0.665,0.495) -- (0.55,0.495) -- (0.55, 0.66);
        \draw[thick, white, ->] (0.75,0.8) -- (0.66,0.68);
        \node at (0.8,0.88) {\color{white}{\large TES area}};
        \draw[thick, white, ->] (0.3,0.83) -- (0.48,0.73);
        \node at (0.3,0.88) {\color{white}{\large SiN$_{\mathrm{x}}$} beams};
    \end{scope}
\end{scope}
\end{tikzpicture}
\caption{\label{fig:TES70} \SI{70}{\micro\meter}}
\end{subfigure}
\begin{subfigure}{\columnwidth}
\begin{tikzpicture}
\begin{scope}
    \node[anchor=south west,inner sep=0] (image) at (0,0) {\includegraphics[width=\textwidth,angle=0]{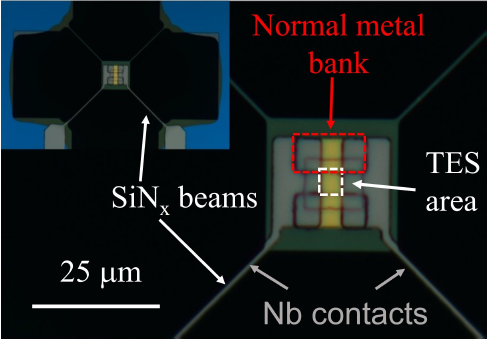}};
    \begin{scope}[x={(image.south east)},y={(image.north west)}]
    \end{scope}
\end{scope}
\end{tikzpicture}
\caption{\label{fig:TES5} \SI{5}{\micro\meter}}
\end{subfigure}
\caption{\label{fig:TESpics} Photos of the (a) largest and (b) smallest device tested here. The TES area, the region used to define the bilayer side length, is indicated by the dashed white square in both photos. The SiN$_{\mathrm{x}}$ legs supporting the membrane are also labelled. Normal metal banks and superconducting contacts are indicated in the close-up photo of (b).}
\end{figure}

The measurements were made on a set of Mo/Au bilayer TESs with Nb wiring that were specifically fabricated to explore a range of device geometries with TES areas varying from $5\times 5 \, {\rm to} \, 70\times 70\,{\rm \mu m^2}$.
Unless otherwise noted, in all devices tested the bilayer was thermally isolated from the substrate by suspending it on a 200\,nm thick SiN island in a well on four 1.4$\upmu$m wide and 50\,$\upmu$m long beams (or `legs').
The thermal conductance between the TES and the substrate was designed to be $G_\text{b}\sim$2\,pW/K. Optical micrographs of representative devices are shown in Fig. \ref{fig:TESpics}.

The TESs were processed on Si wafers coated with a membrane layer comprising 200\,nm thick LPCVD SiN grown on top of a 50\,nm thermal thick SiO$_2$ etch-stop layer.
Reactive-ion etching was first used to define the device wells, islands and support beams in the front- and backside membrane layers.
This was followed by deposition of the metal layers, in each case by DC magnetron sputtering. The Mo/Au bilayers were deposited first and patterned using a two-stage wet etch process.
120\,nm of Au was deposited on top of 40\,nm of Mo, targeting $T_{\crit}$\,$\sim$\,180--200\,mK. 
Next a lift-off process was used to pattern additional 200\,nm-thick Au normal metal features, such as banks and bars, on top of the bilayers.
A lift-off process was then used to pattern a 200\,nm thick Nb wiring layer.
Finally, the device membranes were released from the substrate by using  deep reactive ion etching (DRIE) to remove the underlying Si. This etch was carried out from the back-side of the wafer and terminated on the SiO$_2$ etch stop layer.

Images of the largest and smallest TES tested are shown in Figure \ref{fig:TESpics}.
The white dotted lines in both images indicate the nominal TES dimension and we identify the TESs in terms of the length of this region $L$.
Figure \ref{fig:TES5} highlights the key features of a \SI{5}{\micro\meter} TES.
The red dotted lines show the outline of one of the longitudinal metal banks. Nb leads and two of the ${\rm SiN_x}$ beams are also indicated.

\begin{figure}
\begin{tikzpicture}
\begin{scope}
    \node[anchor=south west,inner sep=0] (image) at (0,0) {\includegraphics[width=8.6cm,angle=0]{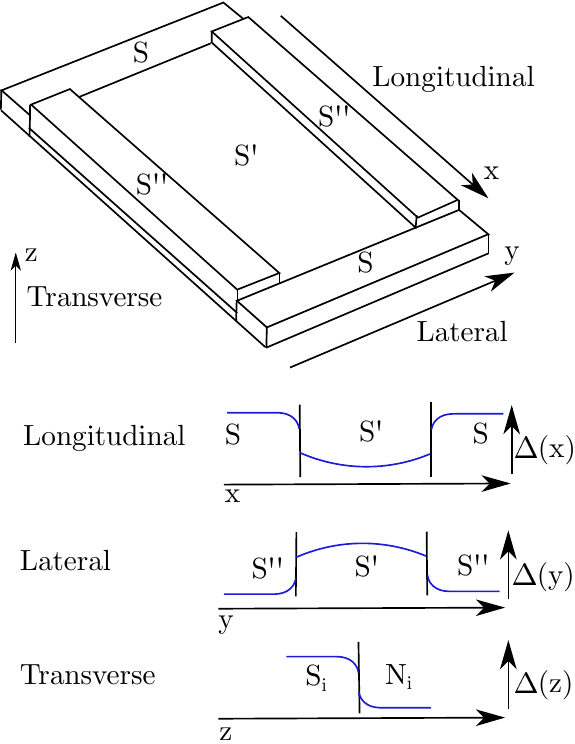}};
    \begin{scope}[x={(image.south east)},y={(image.north west)}]
       \draw[black, dashed, thick] (0.2, 0.83) -- (0.38, 0.89) -- (0.44, 0.85) -- (0.245, 0.79);
       \draw[black, dashed, thick] (0.44, 0.85) -- (0.44, 0.84) -- (0.255, 0.784);
       \draw[black, dashed, thick] (0.6, 0.79) -- (0.41, 0.73) -- (0.465, 0.69) -- (0.66, 0.75);
       \draw[black, dashed, thick] (0.66, 0.75) -- (0.66, 0.74) -- (0.465, 0.682) -- (0.465, 0.69);
       \draw[black, dashed, thick] (0.465, 0.682) -- (0.41, 0.723) -- (0.41, 0.73)  ;
        \draw[black,<->] (0.522,0.45) -- (0.75,0.45) node[above, xshift=-27pt] {$L$};
        \draw[black,<->] (0.518,0.28) -- (0.74,0.28) node[above, xshift=-27pt] {$W$};
    \end{scope}
\end{scope}
\end{tikzpicture}
\caption{\label{fig:proxgeom} Diagram of the TES structure with leads and banks illustrating the proximity effects involved. The underlying physics of the effect is the same regardless of its direction. The superconducting leads are labelled $\Sl$, the thin film TES bilayer is $\Sp$ and the side banks are $\Spp$. The dashed lines show normal metal bars, included on some of the devices, which are the same thickness as the side banks. For these TESs, the regions $\Sp$ and $\Spp$ are composed of a superconductor $\Sl_i$ of uniform thickness and a normal metal $\mathrm{N}_i$, whose thickness differs between $\Sp$ and $\Spp$. The variations of the order parameter, a measure of the superconductivity, are indicated underneath the diagram as a function of position in the $x$, $y$ and $z$ directions.}
\end{figure}
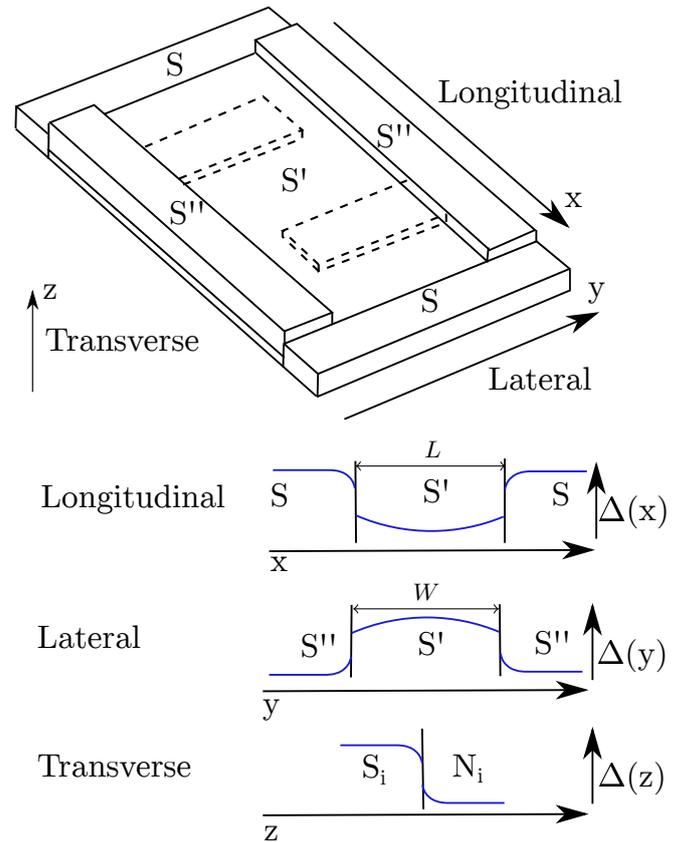

Figure \ref{fig:proxgeom} shows the general geometry and  notation used to describe the components of the TESs.
We assume that current flows in the longitudinal ($x$) direction between the two superconducting electrodes, $\Sl$.
The structures labelled $\Spp$  are the 200\,nm normal metal banks deposited across the longitudinal edge of the bilayers.
These strongly suppress the superconductivity of the underlying bilayer, giving a clean boundary between normal and superconducting regions at the bank edge; in this way irregularities associated with the edge of the bilayer are avoided.
The area of bilayer $\Sp$ left uncovered by banks and wiring was square in all cases and its side-length is used to here define the size of the TES.
In addition, for some TESs partial lateral normal-metal bars were added in the same deposition step as the longitudinal banks.
Two such bars are indicated by the dotted lines in the diagram in Fig.~\ref{fig:proxgeom}.
The addition of the normal metal features on the longitudinal edges also creates a lateral proximity effect. The lower plots sketch typical representative spatial variations of the superconducting order parameter in all three dimensions \cite{Kozorezov2011}.

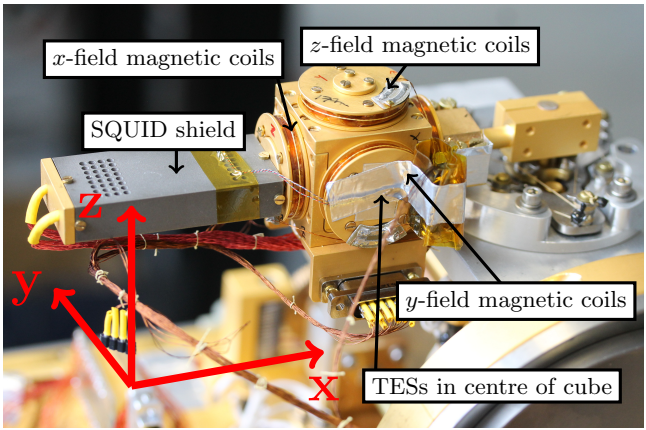
\begin{figure}
\resizebox{\columnwidth}{!}{
\begin{tikzpicture}
\begin{scope}
    \node[anchor=south west,inner sep=0] (image) at (0,0) {\includegraphics[width=\columnwidth,angle=0]{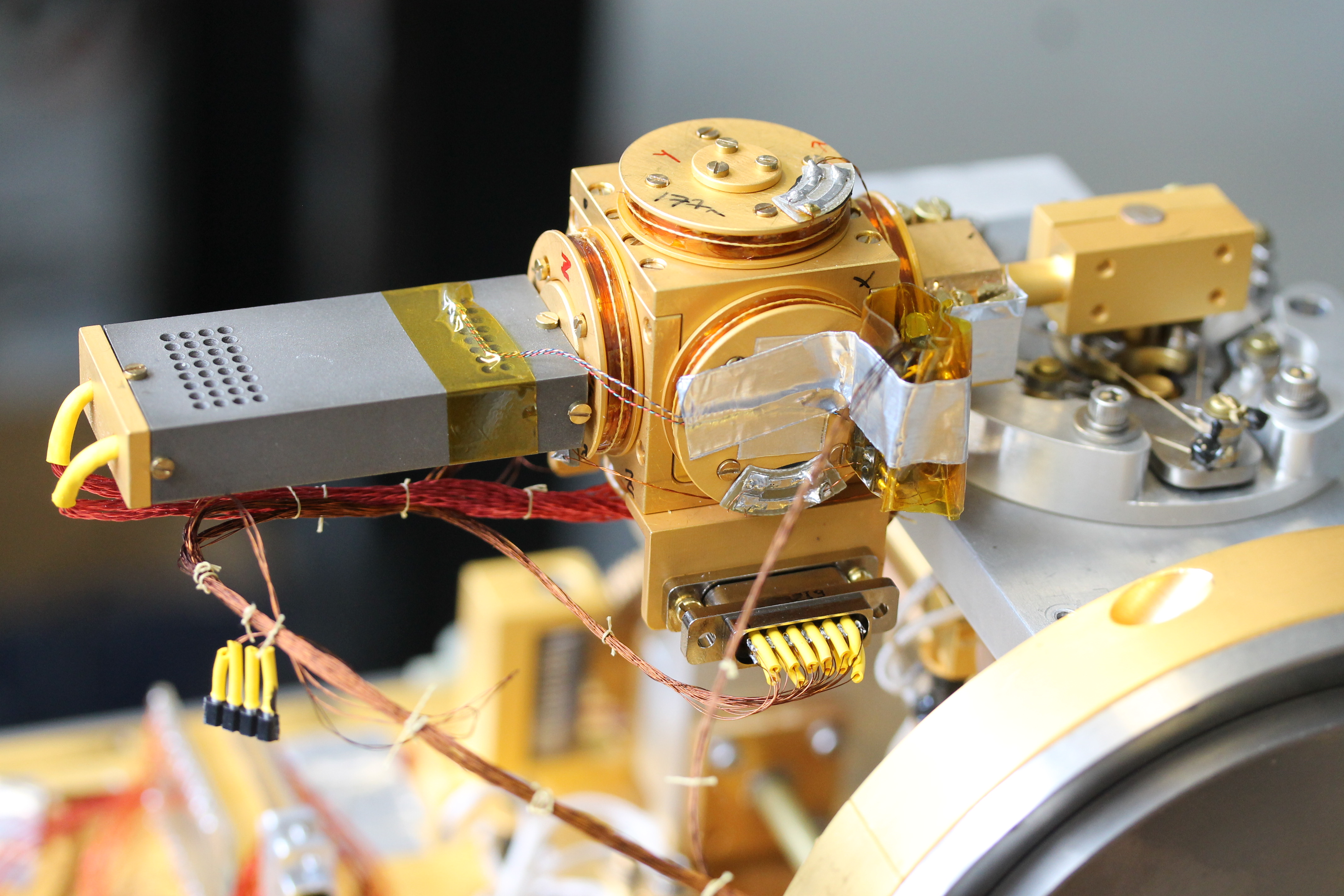}};
    \begin{scope}[x={(image.south east)},y={(image.north west)}]
        \draw[red,->, line width=1.0mm] (0.2,0.1) -- (0.2,0.525) node[left,xshift=-6pt,font=\huge] {\bf z};
        \draw[red,->, line width=1.0mm] (0.2,0.1) -- (0.5,0.18) node[below,yshift=-5pt,font=\huge] {\bf x};
        \draw[red,->, line width=1.0mm] (0.2,0.1) -- (0.08,0.33) node[left,font=\huge] {\bf y};
        \draw[very thick, black, ->] (0.27,0.7) -- (0.27,0.6);
        \node[fill=white,draw=black, very thick] at (0.25,0.7) {SQUID shield};
         \draw[very thick, black, ->] (0.65,0.9) -- (0.6,0.8);
        \node[fill=white,draw=black, very thick] at (0.65,0.9) {$z$-field magnetic coils};
         \draw[very thick, black, ->] (0.42,0.87) -- (0.45,0.7);
        \node[fill=white,draw=black, very thick] at (0.25,0.87) {$x$-field magnetic coils};
         \draw[very thick, black, ->] (0.8,0.3) -- (0.63,0.6);
        \node[fill=white,draw=black, very thick] at (0.8,0.3) {$y$-field magnetic coils};
        \draw[very thick, black, ->] (0.58,0.1) -- (0.59,0.55) node[fill=white,draw=black,xshift=-5pt,pos=0,right] {TESs in centre of cube};
    \end{scope}
\end{scope}
\end{tikzpicture}
}
\caption{\label{fig:expsetup}
TES box mounted on the lowest temperature stage of an ADR. The SQUIDs are mounted in the protective niobium shield seen on the left of the figure.
The field coils  can be seen on the sides of the box.
The coordinate axes, shown in red, are used to describe the direction of the applied magnetic field.
The TESs were mounted with the plane of the film perpendicular the $z$ axis and the direction of current flow parallel to the $y$ axis.}
\end{figure}

The TESs were mounted in the box shown in Figure \ref{fig:expsetup} for testing.
This incorporated a three-axis Helmholtz arrangement, with three pairs of current-carrying coils mounted on the sides of a cube.
The mechanical design of the assembly ensured that the TESs were located at the centre of the coils, with the plane of the TESs normal to $B_z$ and the longitudinal axis parallel to $B_x$.  
This magnetic coil arrangement is not a true Helmholtz 3-axis arrangement, since each pair of coils is separated by a distance equal to their diameter instead of their radius.
However, the expected field in the \SI{5}{\milli\meter} by \SI{5}{\milli\meter} region in the  centre of the cube occupied by the devices was simulated and shown to be uniform to within 2-3\%.
The magnitude of the applied field in each direction was calibrated using a Lakeshore cryogenic Hall probe \cite{Harwin2019}.

The TESs were read-out using superconducting quantum interference devices (SQUIDs).
The SQUIDs were mounted close to the TESs but surrounded by a machined Nb shield to ensure that their operation was not affected by the operation of the magnetic field system.
At the maximum magnetic field used, the change in the measured SQUID current corresponded to a variation in TES current of \SI{30}{\nano\ampere}.
Since the TES currents are of the order of \SI{}{\micro\ampere}, this effect was not significant.
This Nb shield can be seen in the left hand side of Figure \ref{fig:expsetup}.

The TESs and SQUID electronics were cooled to operating temperature using a two-stage adiabatic demagnetisation refrigerator (ADR) launched from a pulse-tube cooled 4\,K stage. 
The base temperature of the system was $\sim 65\,{\mathrm {mK}}$.

\section{Results}\label{Sec:Results}

\subsection{The effect of magnetic field on critical current}\label{Sec:Effect_of_field_on_Jc}

Here we describe measurements of the TES critical current $I_{\crit}$ with magnetic field $B_z$ normal to the plane of the TESs.
For these measurements the bath temperature was regulated at a temperature close to $T_{\crit,i}$ to better than $\pm 0.1\,\mathrm{mK}$.
For clarity here and in what follows, we identify individual TES characteristics  particularly $T_{\crit}$ with an additional subscript e.g. $T_{\crit,i}$ to indicate that measurements apply to a specific device. Our measured  $T_{\crit,i}$ varied with TES size. 
For each applied field, and each TES the  bias voltage $V_b$ was swept symmetrically about $V_b = 0$ and $I_{\crit}$ determined from the device current at the first onset of resistance for both positive and negative bias directions.
The voltage sweep was carried out at a sufficiently low rate of 0.3\,Hz to avoid thermal hysteresis, similar to Smith et al. \cite{Smith2013}.
Critical current at zero applied field was of order \SI{10}{\micro\ampere} for all devices.

Assuming that the supercurrent density is uniform, a magnetic field $B_z$ is expected 
to modulate the TES critical current such that \cite{Smith2013}
\begin{equation}\label{eq:sincIcB}
	I_{\crit} (\Phi_z) = I_{\crit}(0) \left| \mathrm{sinc} (\pi \Phi_z/\Phi_0) \right|.
\end{equation}
Here $I_{\crit}(0)$ is the critical current at zero field, $\Phi_z=B_z A_{\mathrm{eff}}$ is the magnetic flux through the plane of the TES and $\Phi_0$ is the magnetic flux quantum.
$A_{\mathrm{eff}}$ is an effective TES area defined further below.
The variation of $I_{\crit} (\Phi)$  is then the same as the Fraunhofer diffraction pattern.
Equation~\ref{eq:sincIcB} suggests that a smaller TES is less sensitive to a normal magnetic field $B_z$.

It is necessary to assume a value of $A_\text{eff}$ to convert the known applied field to the flux $\Phi_z$.
For a TES with order parameter modified by the proximity effect as shown in Fig.~\ref{fig:proxgeom}, it is not obvious that the effective area $A_{\mathrm{eff}}$ for the flux calculation is the same as the geometric bilayer area $A$, as indicated in Fig.~\ref{fig:TES5}.
In particular, it might be expected that the contribution to $A_\text{eff}$ from a particular area element will scale with the value of the superconducting order parameter at that point.
The approach we take here is to to instead determine $A_\text{eff}$ for a given device as the area giving the best account of the central peak  of the measured supercurrent oscillations.
From the effective area we can define an effective side length $L_{\mathrm{eff}} = \sqrt{A_{\mathrm{eff}}}$.

\begin{figure}
\includegraphics[width=\columnwidth]{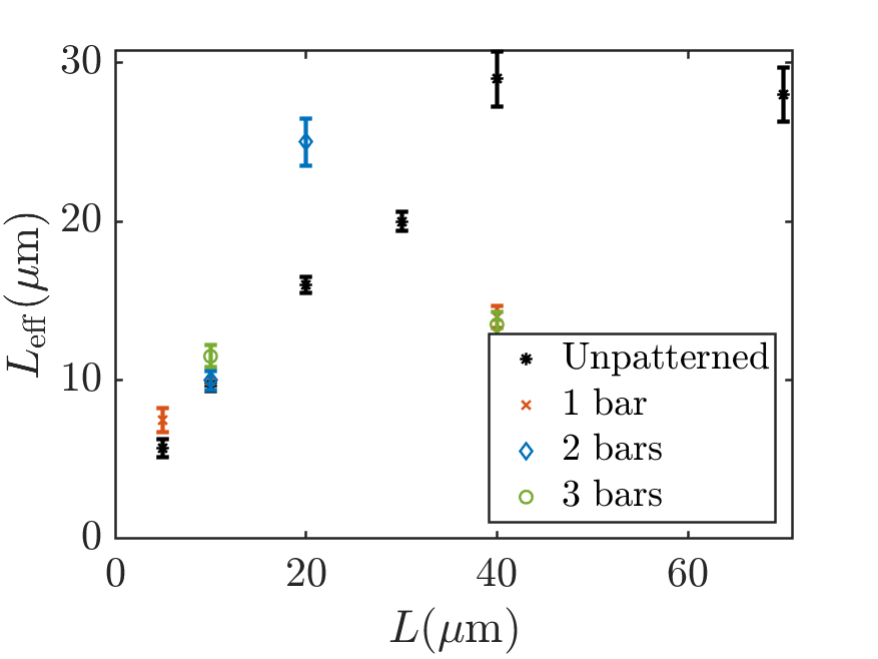}
\caption{\label{fig:SIZEeffL} Effective length, corresponding to the effective area determining by fitting (\ref{eq:sincIcB}) to the $I_\text{c}$ vs B data, as a function of true length.
Black asterisks show data for unpatterned devices, red crosses show devices with a single partial bar, blue diamonds show devices with two partial bars and green circles show devices with three partial bars.}
\end{figure}

Figure \ref{fig:SIZEeffL} shows the effective length calculated in this manner plotted as a function of true length for all of the devices tested in this study.
For unpatterned devices with no bars the effective length is always smaller than, but comparable with, the actual length, with greater deviation seen for larger devices.
The behaviour for devices with bars is more complicated and will be discussed in subsequent sections.

\begin{figure}[!h]
\includegraphics[width=\columnwidth]{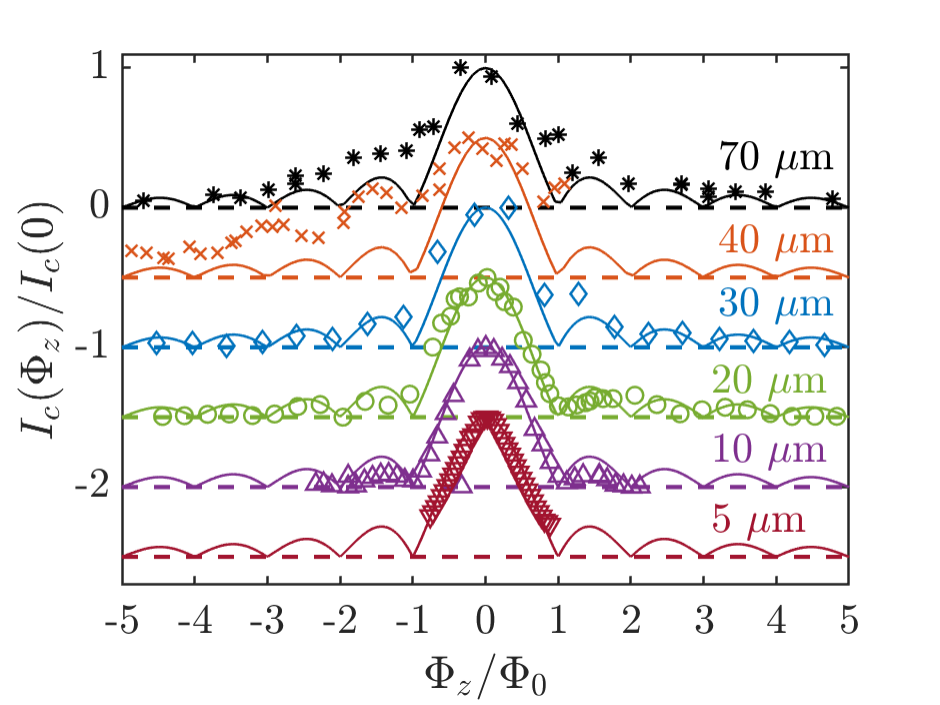}
\caption{\label{fig:SIZEIcBdata} Critical current as a function of magnetic flux, calculated using the effective area that gives the best fit to the oscillations in $I_{\crit}$, for a series of unpatterned TESs with different bilayer side lengths.
Measurements for each TES are plotted with a vertical offset of 0.5 between datasets where the coloured-matched dashed lines indicate the y-axis zero in each case. 
Each set of measurements was taken just below the transition temperature of the TES under test.
All data sets have been normalised to $I_{\crit}$ at zero flux.
The solid lines show the prediction of Eq.~\eqref{eq:sincIcB}.}
\end{figure}

\subsubsection{$I_{\crit}(\Phi)$ for unpatterned devices}\label{Sec:unpatterned_Jc_with_field}

Figure \ref{fig:SIZEIcBdata} shows the measured normalized critical currents $I_{\crit}(\Phi)/I_{\crit}(0)$ as a function of magnetic flux for a series of unpatterned TESs with different bilayer side lengths in the range 5 to \SI{70}{\micro\meter}.
The solid lines show the fitted dependence of $I_{\crit}(\Phi)$ as given by Eq.~\eqref{eq:sincIcB}.
Measurements for each TES are plotted with a vertical offset of 0.5 between datasets where the coloured-matched dashed lines indicate the y-axis zero in each case. 
For the smaller devices, \SI{20}{\micro\meter} side length and below, very good agreement is shown with the prediction, whilst  the larger devices are less well described.
In particular, the larger devices do not show well-defined minima in $I_{\crit}$ when the flux is an integer of flux quanta.

\subsubsection{Effect of partial lateral bars on $I_{\crit}(\Phi)$}
\label{Sec:Effect_of_bars_on_Jc_with_field}

\begin{figure}
\includegraphics[width=8.6cm]{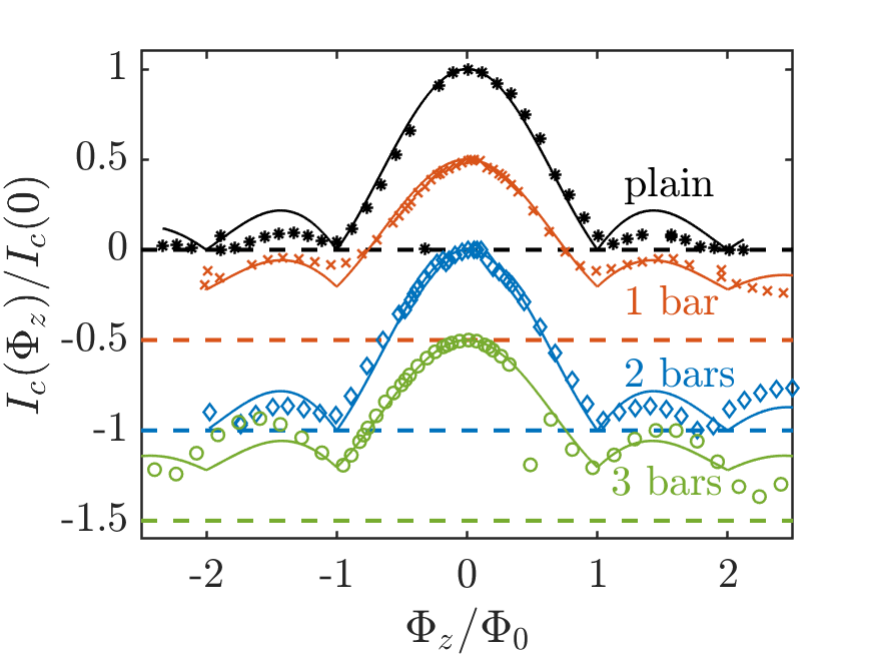}
\caption{\label{fig:IcB10} Critical current as a function of magnetic flux through the bilayer for a series of \SI{10}{\micro\meter} devices patterned with different numbers of normal metal bars. All data sets have been normalised to $I_{\crit}$ at zero flux, and each set of measurements was taken just below the transition temperature of the TES under test. The data are plotted with a vertical offset of 0.5 between datasets. The solid lines for the unpatterned TES and the TES with two bars show the prediction of Eq.~\eqref{eq:sincIcB} whilst the solid lines for TESs with one and three bars have been fitted using Eq.~\eqref{eq:sincIcBmod}. Again the colour-matched horizontal dashed lines indicate the y-axis for the offset data sets. }
\end{figure}

Figure \ref{fig:IcB10} shows the measured the critical current as a function of magnetic flux for TESs a series of \SI{10}{\micro\meter} bilayers with increasing numbers of bars.  Again the colour-matched horizontal dashed lines indicate the y-axis for the offset data sets. 
The currents have been normalised to $I_{\crit}$ at zero flux and the magnetic flux has been calculated using the effective area that gave the best fit to the periodic structure in critical current.
At first sight, Fig.~\ref{fig:IcB10} shows the expected Fraunhofer-like dependence of $I_{\crit}(\Phi)$ with $\Phi$ for all of these \SI{10}{\micro\meter} TESs.
However, we find a clear difference between behaviour of $I_{\crit}(\Phi)$ for between even and odd numbers of bars.
For even numbers of bars  $I_{\crit}(m \Phi_0) <0.02 I_{\crit}(0)$ for integer $m$, as expected from Eq. \eqref{eq:sincIcB}. 
For odd numbers of bars we instead find $I_{\crit}(m \Phi_0) \simeq 0.4 I_{\crit}(0)$ for integer $m$, i.e. although they are local minima in the critical current, the points $\phi_z = m \phi_0$ for integer $m$ are not the expected nulls.

Empirically we found that the behaviour of the critical current for odd numbers of bars could be well described by modifying Eq.~\eqref{eq:sincIcB} with the addition of a quadratic term:
\begin{equation}\label{eq:sincIcBmod}
	I_{\crit} (\Phi) = I_{\crit}(0) \left( a_1 \left| \mathrm{sinc} (f(\Phi)) \right| + (1-a_1) - a_2 (f(\Phi))^2 \right), 
\end{equation}
where $f(\Phi) = \pi \Phi_z/\Phi_0$ and $a_j$ are parameters of the model.
The solid lines shown in Fig.~\ref{fig:IcB10}~(a) for 1 and 3 bars use $a_1$=0.7 and $a_2$=0.005. 
Despite the evidently different behaviour of  $I_{\crit} (\Phi)$ with bar number, all four TESs have similar effective lengths, as seen in Fig.~\ref{fig:SIZEeffL}.

The dependence of $I_{\crit}(\Phi)$ on number of bars may arise from the geometry of the supercurrent flow in the different cases.
Measurements on identical TESs with lateral bars that completely crossed the $\Sp$ region have been carried out by Harwin\,\,\cite{Harwin2020}.
These showed that the bar/bilayer structure itself remained in the normal state to temperatures at least to $80\,{\rm mK}$, i.e. well below the temperatures of interest here, and with zero measurable supercurrent, despite the proximity of the bilayer.
This implies negligible supercurrent current flow across the bars themselves, so partial bars should result in a meandering supercurrent pattern.

\begin{figure}
\begin{subfigure}{\columnwidth}
\includegraphics[width=\textwidth]{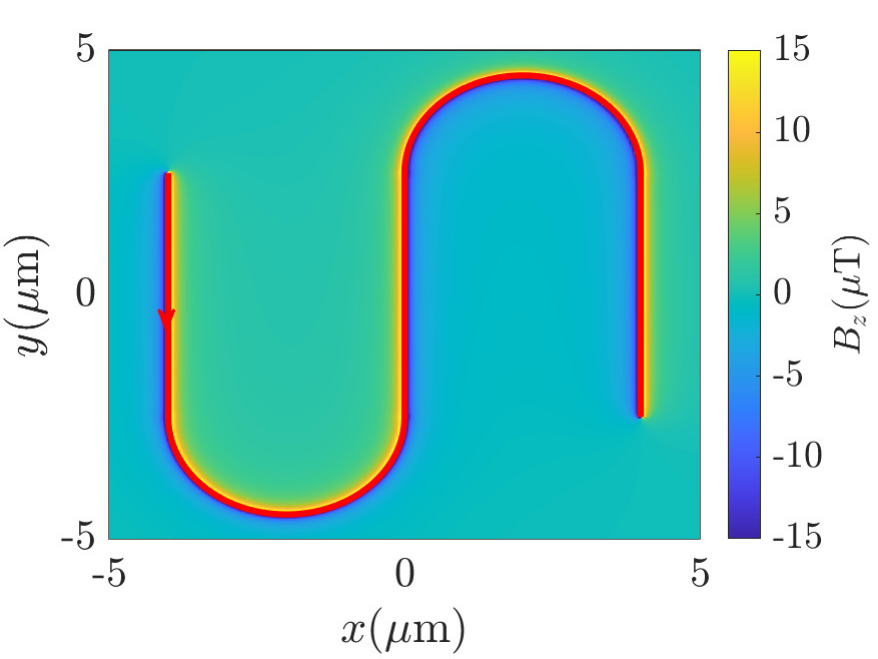}
\caption{\label{fig:Bmeander2} Two bars}
\end{subfigure}
\begin{subfigure}{\columnwidth}
\includegraphics[width=\textwidth]{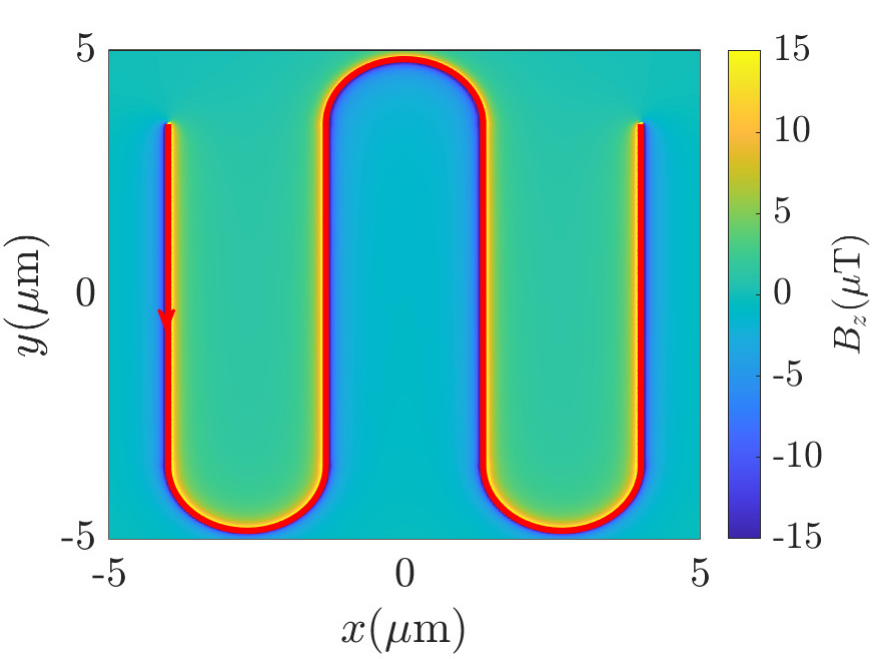} \\
\caption{\label{fig:Bmeander3} Three bars}
\end{subfigure}
\caption{\label{fig:Bmeander} Calculated magnitude of the component of magnetic field perpendicular to the TES thin film, over the surface of the film, produced by a current of   $5\,{\rm\upmu A}$  following the path shown in red. (a) and (b) illustrate the effects of differing numbers of normal metal bars, as the normal metal bars determine the pattern of current flow.}
\end{figure}

The self-magnetic field generated in the TES by a meandering current can be understood using a simplified Biot-Savart law calculation.
Figure~\ref{fig:Bmeander} shows $B_z$  produced by a current of $20\,{\rm\upmu A}$  following the path shown in red.
Figure~\ref{fig:Bmeander2} shows the case with two normal metal bars and Fig.~\ref{fig:Bmeander3} the case with three normal metal bars.
Critically, it can be seen that sign of the self-field alternates between adjacent current  loops.

The self-flux is the sum of the flux in each of these loops and adds to the applied flux, modifying the behaviour of $I_\text{c}$.
With an even number of bars and therefore loops, we expect the sums of fluxes in adjacent pairs of loops to cancel; hence the self-field correction will be negligible.
For an odd number of bars there is one unmatched loop, leading to a finite self-flux correction.
Further, the scale of the correction should decrease as the applied B-field is increased and the self-field becomes less significant.
This behaviour is consistent with the data.

\begin{figure}
\includegraphics[width=8.6cm]{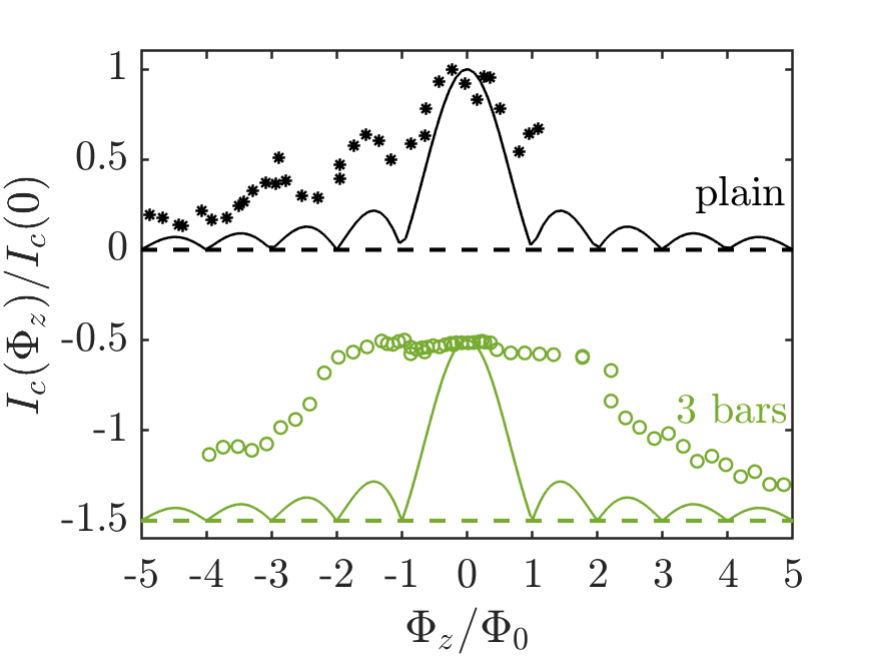}
\caption{\label{fig:IcB40} Critical current as a function of magnetic flux through the bilayer for two devices with 
$40\,{\rm\upmu m}$
side-length bilayers, one unpatterned and one with three partial normal metal bars. Both data sets have been normalised to $I_{\crit}$ at zero flux, and each set of measurements was taken just below the transition temperature of the TES under test. The data are plotted with a vertical offset between datasets of 1.5. The solid lines show the prediction of Eq.~\eqref{eq:sincIcB}}
\end{figure}

Figure~\ref{fig:IcB40} shows the measured effect of adding three partial normal metal bars to a larger TES with a side length of
$40\,{\rm\upmu m}$.
The Fraunhofer pattern in $I_{\crit}(\Phi)$ is evident for the device without bars, although the expected minima with $\Phi$ are not well defined.
In the device with three bars the Fraunhofer pattern is absent and $I_c$ is almost  independent of $\Phi_z$ up to $3\Phi_0$.
We explore this effect further in Sec.~\ref{Sec:Effect_of_field_on_Resp}.

\subsection{Temperature dependence of $I_{\crit}$ in zero field}\label{Sec:Effect_of_temperature_on_Jc}

\begin{figure}
\begin{subfigure}{\columnwidth}
\includegraphics[width=\textwidth]{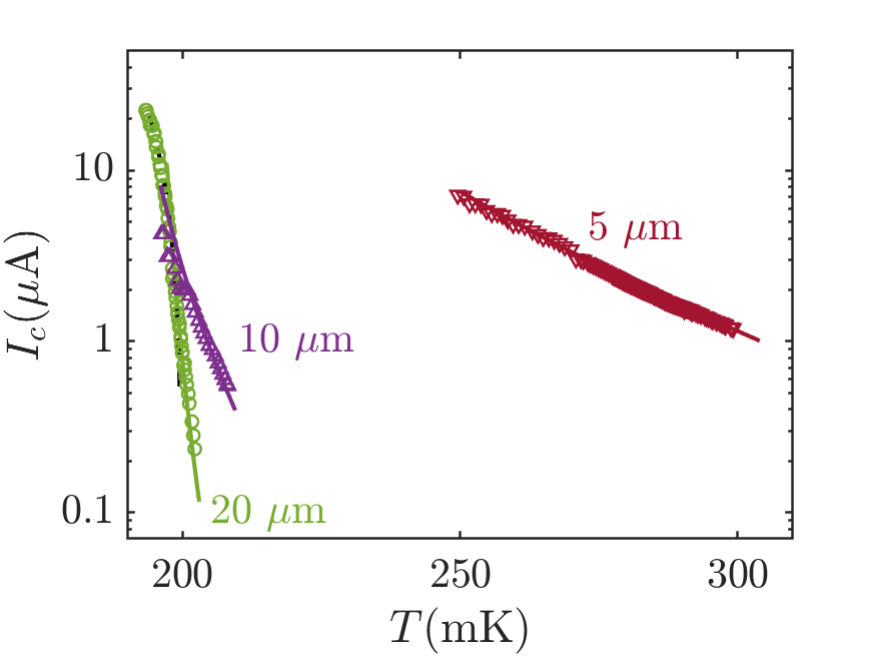}
\caption{\label{fig:IcTlarger}}
\end{subfigure}
\begin{subfigure}{\columnwidth}
\includegraphics[width=\textwidth]{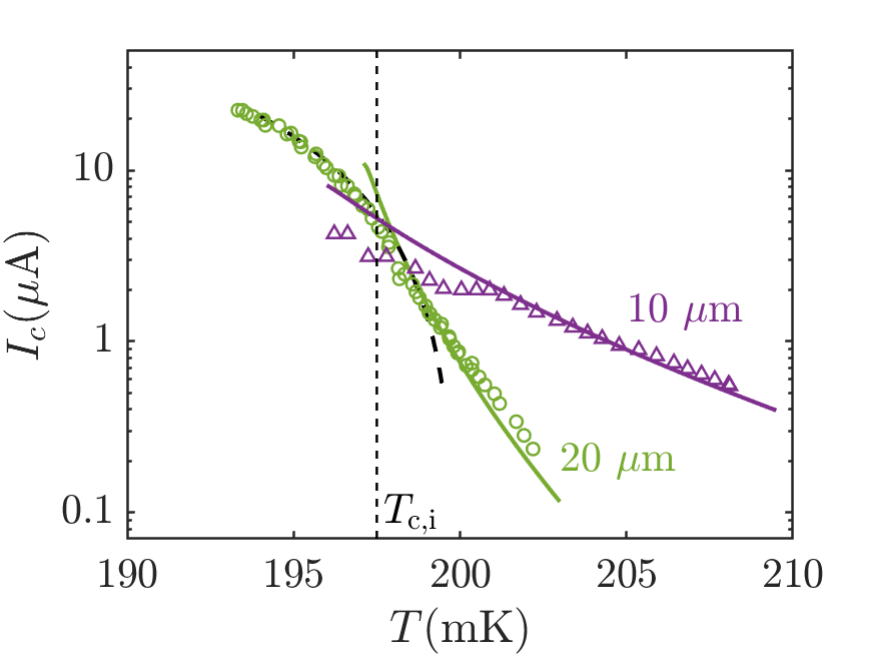} \\
\caption{\label{fig:IcTsmaller}}
\end{subfigure}
\caption{\label{fig:TESIcT} $I_{\crit}(T)$ for three TESs of different bilayer side lengths. All measurements were taken with zero applied field. (a) shows all data and fits, (b) provides more detail of the region near $T_{\crit,i}$. The solid lines show the expected current for a SS'S weak-link for $T>T_{\crit,i}$ for all three data sets, whilst the dashed lines for the 
$20\,{\rm\upmu m}$
data show the expected Ginzburg-Landau dependence below $T_{\crit,i}$ for the 
$20\,{\rm{\upmu m}}$ TES, the dashed vertical line in the lower plot.} 
\end{figure}

Figure~\ref{fig:IcTlarger} shows the critical currents of three of the smallest unpatterned TESs in zero field as a function of $T_b$.
Figure~\ref{fig:IcTsmaller} shows the same data and fits but expands the region around $T~\sim T_{\crit,i}$.
The black lines show the expected temperature dependence of the current in a SS'S weak-link for $T>T_{\crit,i}$ given by both Golubov \cite{Golubov2004} and Sadleir \cite{Sadleir2010} 
\begin{equation}\label{eq:SSS_weaklink_I}
	I_{\crit}(T,L)\propto \frac{L}{\xi(T)} \exp\left(  \frac{-L}{\xi(T)} \right),
\end{equation}
with $\xi(T)=\xi_i/(T/T_{\crit,i}-1)^{1/2}$. For the three TES  lengths, we find a value  $\xi_i=450\pm20\,{\rm nm}$ and $T_{\crit,i} =197\pm3\, {\rm mK}$.
This  gives an excellent account of the measured supercurrents in this regime. 
Below $T_{\crit,i}$, $I_c$ is well accounted for by the expected Ginzburg-Landau dependence $I_{\crit}(T) \propto (1-T/T_{\crit,i})^{3/2}$.
This dependence is shown by  the dashed line in Fig.~\ref{fig:IcTsmaller} for the $20\,{\rm \upmu m}$ TES. 
Using a free electron model with Fermi velocity $v_{\textrm F}=1.39\times 10^{6}\,{\rm m s^{-1}}$, 
and the measured resistivity for our 120 nm-thick Au films $\rho=1.33\times10^{-8}\,{\rm \Omega m}$, 
we calculate a normal-state electronic diffusion coefficient $D=2.80\times10^{-2}\,{\rm m^2s^{-1}}$. 
Using $\xi_n = \sqrt{\hbar D/2\pi k_{\mathrm{B}}T_{\crit}}$, we  calculate  $\xi_n$ = 450\,nm, in good agreement with the measurement. These observations are consistent with Ref.~\onlinecite{Sadleir2010} and we identify
$\xi_i\equiv \xi_n$. 

\subsection{Effect of field on voltage-biased TESs}\label{Sec:Effect_of_field_on_voltage_biased TESs}


Sadleir {\it et al.}  \cite{Sadleir2011,Sadleir2010} investigated the variation of mesured  $T_{\crit}$ with geometry and field for TESs at temperatures $T > T_{\crit, i}$ where $T_{\crit, i}$ is the temperature corresponding to the mid-point of the resistive transition, $R = R_{\nn}/2$. 
Those measurements were performed on unreleased TESs, for which the thermal conductance to the heat bath is high and TES Joule power heating with current bias $P_{\rm J}=I_{b}^2R$ is relatively unimportant for low bias currents $I_{b}$.
Here we report $T_{\crit, i}$ for TESs with thermal isolation from the substrate under voltage bias using the measured bias power at $R_0 = R_{\nn}/2$ as a function of applied field in order  to determine $T_{\crit, i}(B_z)$.
This approach explicitly explores the operating regime of a TES under voltage bias. 

\begin{figure}
\includegraphics[width=\columnwidth]{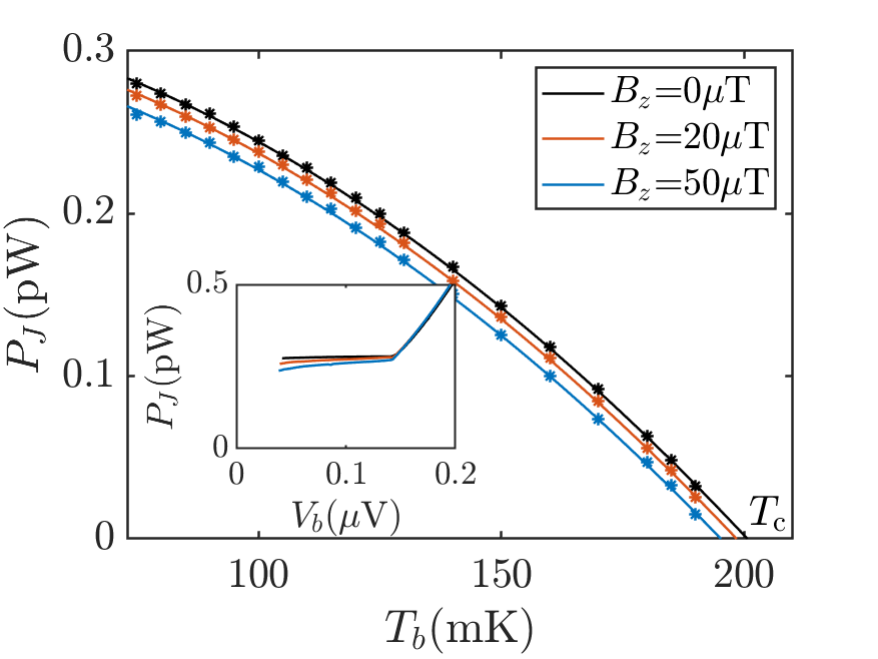}
\caption{\label{fig:SIZEPT} $P_{\rm J} (T_b)$ curves for different applied fields, measured for an unpatterned TES with 
$70\,{\rm\upmu m}$
side length.   $P_{\rm J}$ was determined at $R=0.5R_{\nn}$ for all TESs. The inset shows sample $P(V_b)$ curves at the same applied magnetic fields, all measured at $T_b =$\SI{75}{\milli\kelvin}.}
\end{figure}

Figure~\ref{fig:SIZEPT} shows measurements of the TES Joule power determined at $R =R_{\nn}/2$ for an unpatterned TES with 
$70\,{\rm\upmu m}$
side length. 
These  $P_{\rm J}(T_b)$ curves were parametrized using $P_{\rm J}(T_b) = K (T_{\crit,i}^n -T_b^n)$, assuming $K$ and $n$ do not change with field for a particular TES. 
This procedure was used to determine $T_{\crit}(B)$ for a range of TES lengths.
The inset of Fig.~\ref{fig:SIZEPT} shows example $P_{\rm J}(V_b)$  curves at three different applied magnetic fields for this 
TES,  all measured at fixed $T_b =$\SI{75}{\milli\kelvin}.
The $P_{\rm J}(V_b)$ plot at the maximum field shown here, $B=50\,{\rm \upmu T}$, would correspond to $ \Phi/\Phi_0 \sim 15$, well-beyond the range of data shown in Fig.~\ref{fig:SIZEIcBdata}.
 
\begin{figure}
\begin{subfigure}{\columnwidth}
\includegraphics[width=\textwidth]{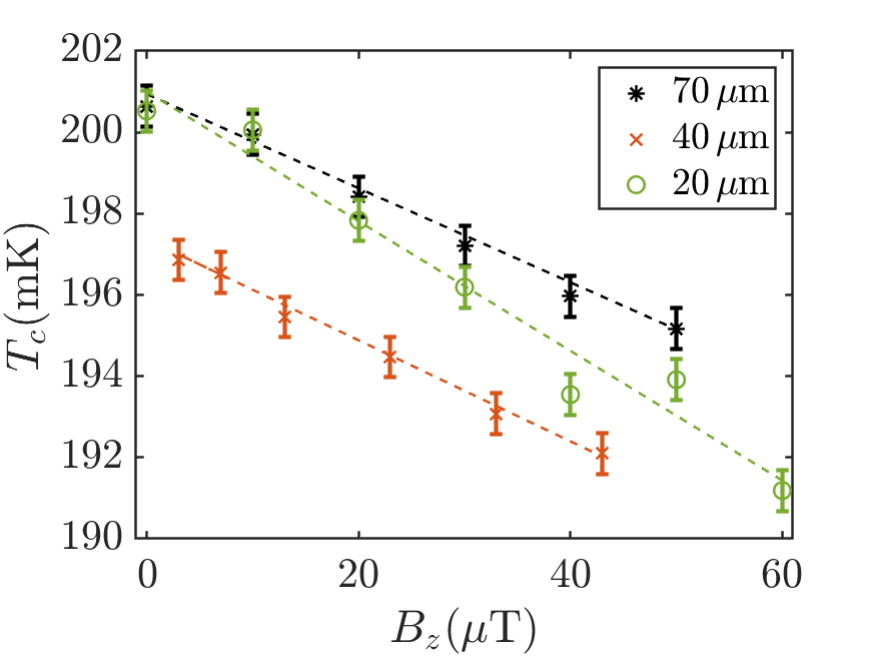}
\caption{\label{fig:TcBlarger}}
\end{subfigure}
\begin{subfigure}{\columnwidth}
\includegraphics[width=\textwidth]{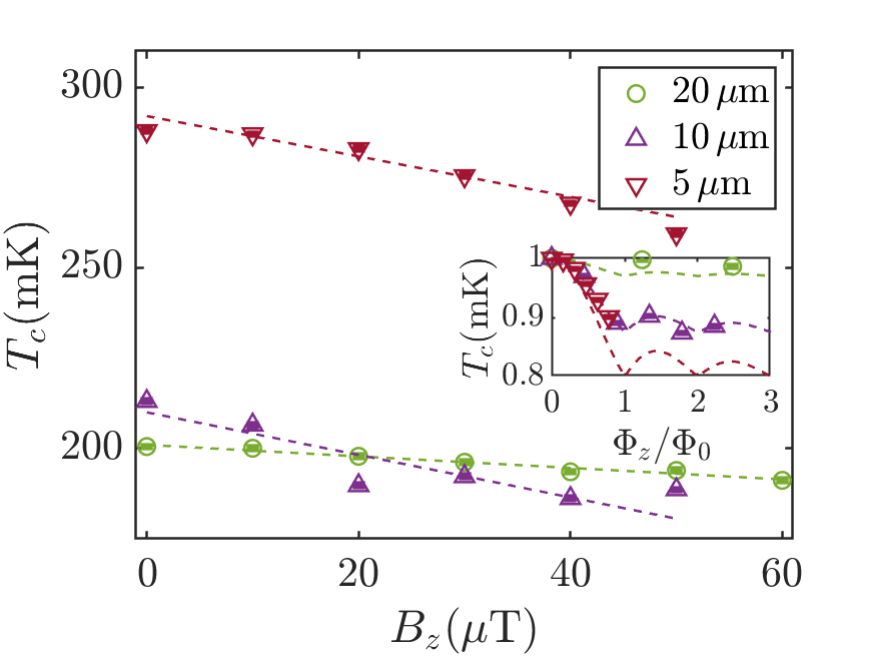} \\
\caption{\label{fig:TcBsmaller}}
\end{subfigure}
\caption{\label{fig:SIZETcB}
Transition temperature as a function of magnetic field for a series of unpatterned TESs with different bilayer side lengths.
Each data point was calculated from a series of $I(V)$ curves at different temperatures.
(a) is for the larger devices whilst (b) is for the smaller ones.
The measurements were taken with the devices under voltage bias.
Both figures show straight-line fits to the data as dashed lines.
The lower figure inset shows $T_{\crit}/T_{\crit}(0)$ as a function of flux, $\Phi_z/\Phi_0$.
The fits in this inset are given by the weighted combination of Eq. \eqref{eq:sincIcB} and a constant, representing the contributions from the superconducting and normal components of the current respectively.}
\end{figure}

Interestingly, although the field-dependent {\textit {maxima}} of the supercurrents shown in Fig.~\ref{fig:SIZEIcBdata} closely-follow the expected Fraunhofer-like dependence, the Joule power for the voltage-biased TES is changed only modestly ($\sim 5\%$) with $B$.
Additionally there is evidence of change in the overall $P_{\rm J}(V)$ with $B$ that would be consistent with reduced $\alpha$ with increasing field.

Figure~\ref{fig:SIZETcB} shows how  $T_{\crit}(B)$ changes with applied field $B_z$.
Figure~\ref{fig:TcBlarger} shows $T_{\crit, i}(B_z)$
for three unpatterned TESs with geometric length 
$>20\,{\rm\upmu m}$. $T_{\crit,i}$ is reduced almost linearly with applied $B_z$.
It is interesting to investigate this observation in terms of flux-flow resistance in a Type II superconductor \cite{Lefloch1999}.
The applied field would  then be identified as the critical field $B_{\crit,i}$ that gives the measured $T_{\crit,i}$ for each TES. 
The gradients of the linear fits to the data, $\partial B/\partial T |_{B\to0}$, with  $B_{\crit} = \phi_0/2\pi\xi^2$,\cite{Lefloch1999} and $\xi (T) = \xi / \sqrt{1-T/T_{\crit}}$ \cite{Likharev1979}, give an estimate of $\xi $ within this model. 
For the 40 and $70\,{\mathrm {\upmu m}}$ TESs the correlation coefficients are high, $R^2>0.99$,
and we find that $\xi = 470 \pm 20$~nm would account for the field dependence of $T_{\crit,i}$.
For the 
$20\,{\rm\upmu m}$ TES 
the same calculation gives $\xi = 520 \pm 20$ nm although the correlation is lower,  $R^2\simeq 0.95$
At temperatures above the intrinsic TES transition temperature we might expect $\xi = \xi_n$, the normal-state correlation length, indeed the derived $\xi_n$ would agree well with our earlier interpretation of the data shown in Fig.~\ref{fig:TESIcT}.

Interpretation of the measurements of $T_{\crit}(B)$ for smaller 5 and $10\,{\rm\upmu m}$ TESs, shown in Fig.~\ref{fig:TcBsmaller},
  in terms of flux-flow resistance is less compelling.  This would be expected if $L<3.49\xi(T)$ \cite{Likharev1979}. 
The fluctuations in $T_{\crit,i}(B)$ appear more closely related to the Fraunhofer patterns of the underlying $I_c$ as shown in the inset of Fig.~\ref{fig:TcBsmaller}.
For the smaller TESs $J_c$ is reduced to close to zero at the expected minima.

\subsubsection{Effect of applied field on the TES bias current}\label{Sec:Effect_B_on_Ibias}

\begin{figure}
\includegraphics[width=8.6cm]{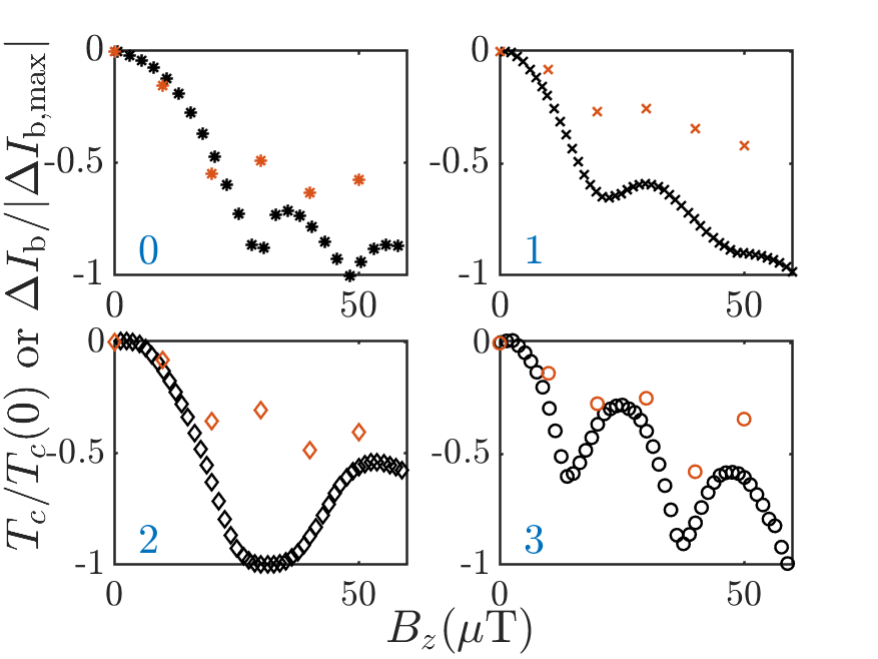}
\caption{\label{fig:osccor} Normalised changes  $\DIb/\Delta I_{\mathrm{bias,max}}$ (black points) and $T_{\crit}/T_{\crit}(0)$ (red points) as a function of  magnetic field $B_z$, for four
$10\,{\rm\upmu m}$
 TESs with different numbers of normal metal bars. The blue number in the lower left corner of each subplot shows the number of partial normal metal bars on the TES.}
\end{figure}

Figure~\ref{fig:osccor} shows the relationship between the oscillations seen in transition temperature and bias current measured with increasing field.
For the four 
$10\,{\rm\upmu m}$ 
side length TESs studied, the change in $T_{\crit}/T_{\crit}(0)$ is plotted on the same axes as the normalised change in current  $\DIb/\DIb(0)$.
There is good agreement between the two sets of oscillations for all four TESs, suggesting that the same effects determine the period of oscillation of both critical current and bias current.
For the TES with two partial bars, the broad minimum in $\DIb$ at around 
$30-40\,{\rm\upmu T}$
corresponds to a maximum in $T_{\crit}$.
The oscillations in $I_{\crit}$ in Fig.~\ref{fig:IcB10} also show a maximum around this region, suggesting that another physical process may be suppressing the maximum in $\Delta I_\text{bias}$.

\subsubsection{Effect of magnetic field on TES responsivity}\label{Sec:Effect_of_field_on_Resp}

\begin{figure}[h!]
\begin{subfigure}{0.5\textwidth}
\includegraphics[width=8.6cm]{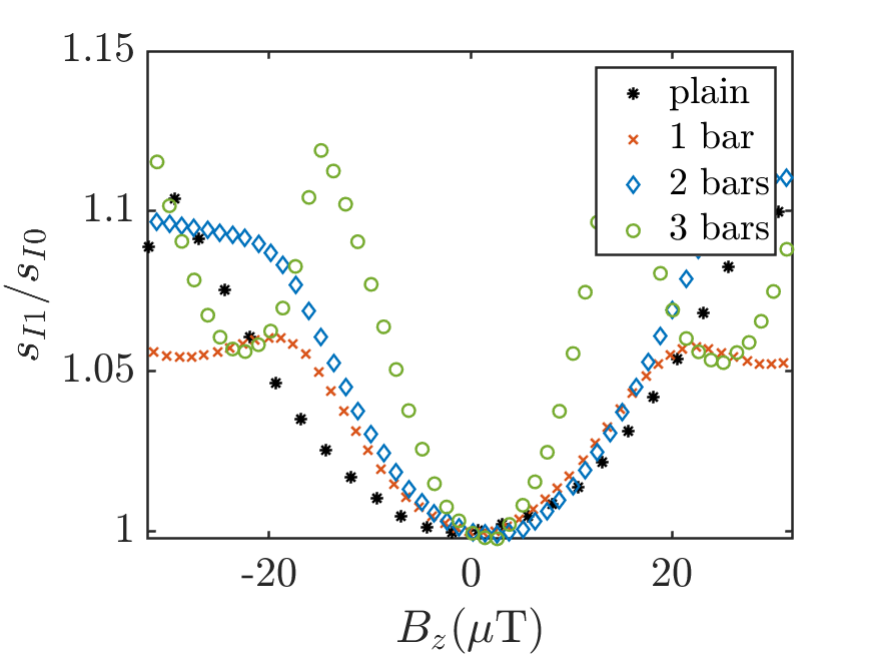}
\caption{\label{fig:IB10} $10\,{\rm\upmu m}$ side length}
\end{subfigure}
\begin{subfigure}{0.5\textwidth}
\begin{tikzpicture}
\begin{scope}
    \node[anchor=south west,inner sep=0] (image) at (0,0) {\includegraphics[width=\textwidth,angle=0]{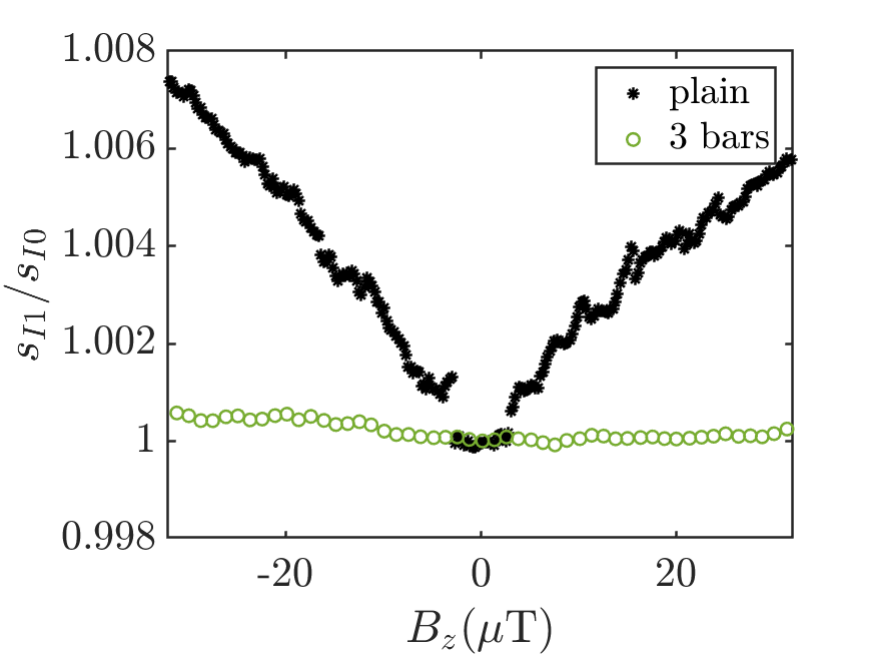}};
    \begin{scope}[x={(image.south east)},y={(image.north west)}]
         \draw[thick, red, ->] (0.55,0.62) -- (0.52,0.375);
          \draw[thick, red, ->] (0.555,0.62) -- (0.575,0.355); 
          \draw[thick, red, ->] (0.65,0.63) -- (0.72,0.61);
         \draw[thick, red, ->] (0.47,0.63) -- (0.38,0.625);             
        \node[align=center,anchor=south] at (0.5575,0.6) {Possible \\ phase slips};
    \end{scope}
\end{scope}
\end{tikzpicture}
\caption{\label{fig:IB40} $40\,{\rm\upmu m}$ side length}
\end{subfigure}
\caption{\label{fig:IBbars}
Change in TES responsivity as a function of magnetic field through the bilayer.
Measurements are for TESs with (a) $10\,{\rm\upmu m}$ side length bilayers and (b) $40\,{\rm\upmu m}$side length bilayers, patterned with different numbers of normal metal bars.  
The bias currents at zero field are (a) plain: 4.52, 1 bar: 3.91, 2 bars: 4.06 and 3 bars: $3.27\,{\rm\upmu A}$  and (b) plain: 2.85 and 
3 bars: $2.32\,{\rm\upmu A}$.
In (b) possible phase slips are indicated for the unpatterned TES.
}
\end{figure}
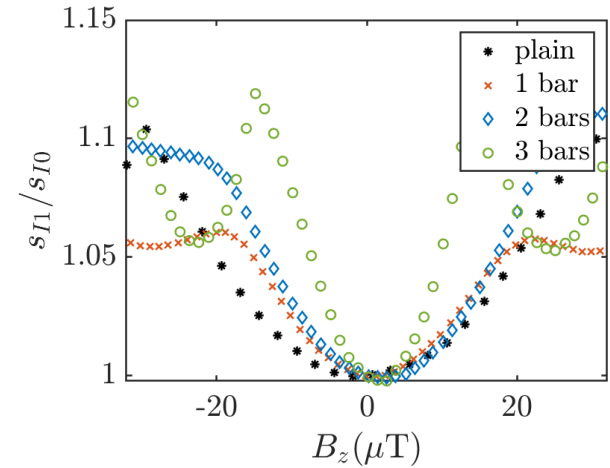
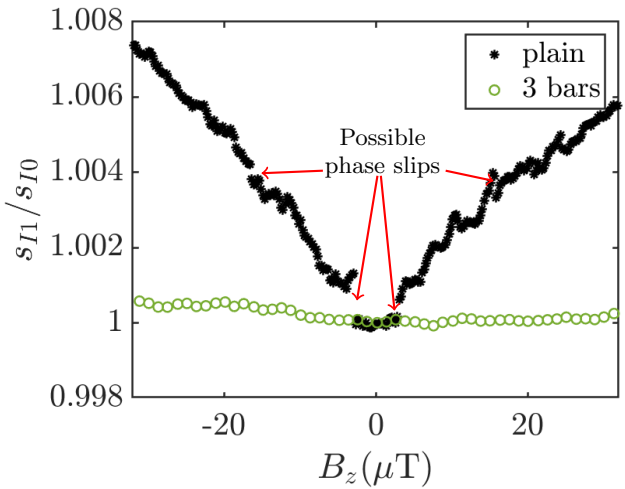

In this section we consider the effect of magnetic fields on TES responsivity when used as a power detector.
To do so it is first useful to consider how the calculation of the responsivity is modified in the presence of a magnetic field.
Magnetic field affects the responsivity of a TES directly through its effect on the bias current (Section~\ref{Sec:Effect_B_on_Ibias}).

TES are usually operated with voltage bias.
However, the most-commonly used TES voltage-bias and readout circuit, as used here, is not an ideal voltage source\,\cite{Irwin2005}.
In the standard circuit analysis, the combination of a low valued bias  resistor $R_b$ (typically $\sim 1\,{\rm m\Omega}$) and any stray resistance $R_{stray}$  are represented by a load resistance $R_L$.
This means that the bias voltage of the TES $V_b$ changes if the TES current changes irrespective of details of the TES behaviour itself.
For simplicity here we assume that the TES electrothermal parameters are near-ideal ($\alpha$ is large, $\beta \to 0$) for all bias points.
With these assumptions and for a TES used as a bolometer, the small-signal  power-to-current responsivity of the TES in a small-signal analysis at an initial bias point is  $s_{I,0}=1/(I_0(R_0-R_L))$ \cite{Irwin2005}.
In this analysis we identify $I_0\equiv I_{b}$ for consistency with the complete small-signal analysis of  Ref.~\onlinecite{Irwin2005}.
If for example a magnetic field changes the bias current so that $I_1=I_0-\Delta I(B)$, the responsivity $s_{I,1}$ is also changed. 
Circuit analysis shows that
\begin{equation}\label{eq:effect_on_responsivity}
	\frac{s_{I,1}}{s_{I,0}}=\frac{I_0(R_0-R_L)}{I_0(R_0-R_L)-2\Delta I(B)R_L}.
\end{equation}
Note that for an ideal voltage bias $s_{I,1}/s_{I,0}=1$ irrespective of $\Delta I(B)$ .
This responsivity is most-relevant for quantifying small-signal power detection in a bolometer, but similar considerations would apply for a calorimeter. 

Figure~\ref{fig:IB10} shows the responsivity as a function of magnetic field for a series of square TESs of 
$10\,{\rm\upmu m}$
 side length.
The oscillations in responsivity are of similar magnitude for all of the TESs.
Figure \ref{fig:IB40} shows the responsivity as a function of magnetic field for two 
$40\,{\rm\upmu m}$
 TESs.
These larger devices display the smallest changes in responsivity with field, and the addition of partial normal metal bars further reduces $s_I$. Figure~\ref{fig:IB40} also shows additional discontinuities at  $\sim\pm3$  and 
$\sim\pm 15\,{\rm\upmu T}$
that maybe indicative of phase-slip behaviour.
These  features were very reproducible as a function of applied field, with finer sampling with field, and even between cool-downs.
The observation that larger TESs may display phase slip behaviour is consistent with the existing theoretical ideas\,\cite{Likharev1979, Golubov2004, Gottardi2021}. These  discontinuities in $I_b(B)$ were not observed in similar larger TESs with additional lateral bars. 

The existing theoretical work does not, however, predict the result that the smallest TESs display the largest changes in current with flux, as these TESs have less magnetic flux threading them, which would imply that any changes in current with flux would be smaller.
This suggests that there may be non-uniformities in the larger TESs, such as vortices, which reduce magnetic flux penetration and hence magnetic field sensitivity.

\begin{figure}
\includegraphics[width=8.6cm]{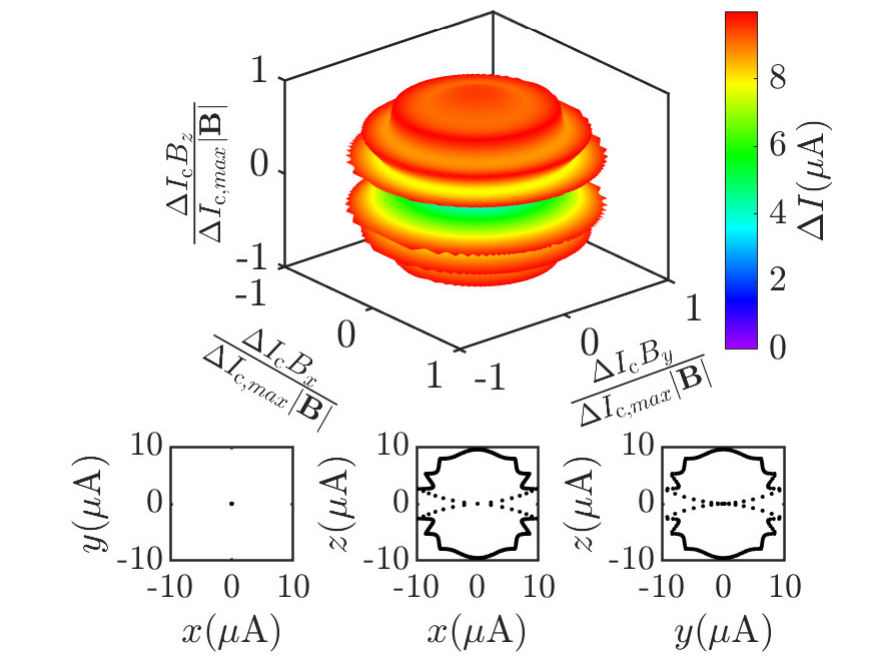}
\caption{\label{fig:responsesurf}
(Top) Change in critical current with direction of the magnetic field applied to a 
$10\times 10\,{\rm\upmu m}$
TES thin film.
In the plot, the direction of the field is scaled by the prediction of the change in critical current from Eq.~\eqref{eq:IBanydirection}, for an applied field of \SI{50}{\micro\tesla}.
The shading corresponds to the magnitude of the response.
The normal of the thin film is in the $z$ direction.
(Bottom) Cuts through the response surface in the $x-y$, $x-z$ and $y-z$ planes. }
\end{figure}

\subsubsection{Directional dependence of bias and critical current on applied field}\label{Sec:Directional_dependence}

To investigate the directional magnetic field dependence of the bias current of the TESs at fixed $V_b$, 
the magnitude of the applied magnetic field was fixed and its direction varied. Using the full 3-axis Helmholtz system, fixing $\vert \mathbf{B} \vert$ defines a sphere of fixed radius, which can be parametrised in terms of  azimuthal and polar angles. 
In order to ensure efficient sampling, the increment of the azimuthal angle was fixed and the increment of the polar angle was varied to sample elements of equal area.

For the general field direction,
\begin{equation}\label{eq:Bvardirec}
	\mathbf{B} =  B_x \hat{\mathbf{x}}+B_y \hat{\mathbf{y}}+ B_z \hat{\mathbf{z}},
\end{equation}
Eq.~\eqref{eq:sincIcB} can be extended so that the critical current is \cite{GrossMarx}
\begin{equation}\label{eq:IBanydirection}
	I_{\crit}(\Phi) = I_{\crit}(0) \prod_{i=x,y,z} \left| \mathrm{sinc} \left( \frac{\pi \Phi_i}{\Phi_0}\right) \right| .
\end{equation}
As $B_x$ is parallel to the direction of current flow, the effect of $\Phi_x$ can be neglected.
Figure \ref{fig:responsesurf} shows the model prediction of the change in critical current with the direction of a \SI{50}{\micro\tesla} applied field, for a 
$10\,{\rm\upmu m}$
side length TES.
The direction of the applied field is scaled by the resulting change in critical current, so for a TES that is equally sensitive to applied fields in all directions, the response surface would be spherical.
The lobed response surface indicates that the thin film is most sensitive to magnetic fields applied perpendicular to the plane of the TES, as expected.
The shading of the surface corresponds to the magnitude of the response.
This model also predicts oscillations in the response, which can be seen both in the response surface and the cuts in the $x-z$ and $y-z$ planes shown underneath. The response is minimal when the applied field is in the $x-y$ plane. The model suggests the general behaviour for $I_b$ of a real TES as a function of field orientation, although experimentally (compare for example Figs.~\ref{fig:SIZEIcBdata} and  \ref{fig:osccor})
although we have already shown that the correlation is imperfect. 

\begin{figure*}[t]
\begin{subfigure}{0.49\textwidth}
\begin{tikzpicture}
\begin{scope}
    \node[anchor=south west,inner sep=0] (image) at (0,0) {\includegraphics[width=\textwidth,angle=0]{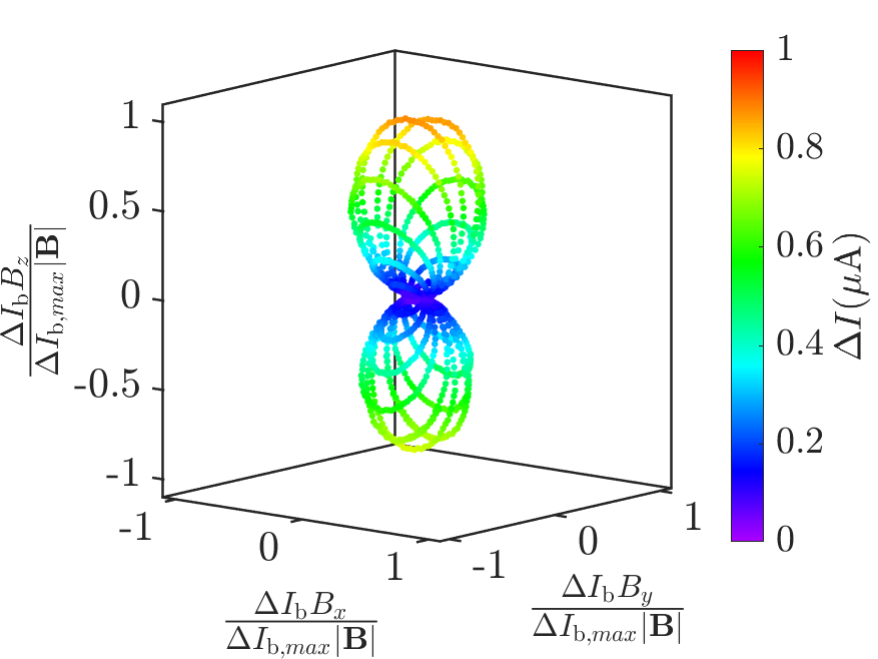}};
    \begin{scope}[x={(image.south east)},y={(image.north west)}]
         \draw[thick, blue] (0.63,0.82) -- (0.63,0.7) -- (0.73, 0.7) -- (0.73, 0.82) -- (0.63, 0.82);
         \node[blue] at (0.68, 0.65) {10$\upmu$m};
    \end{scope}
\end{scope}
\end{tikzpicture}
\caption{\label{fig:dir10p}}
\end{subfigure}
\hfill
\begin{subfigure}{0.49\textwidth}
\begin{tikzpicture}
\begin{scope}
    \node[anchor=south west,inner sep=0] (image) at (0,0) {\includegraphics[width=\textwidth,angle=0]{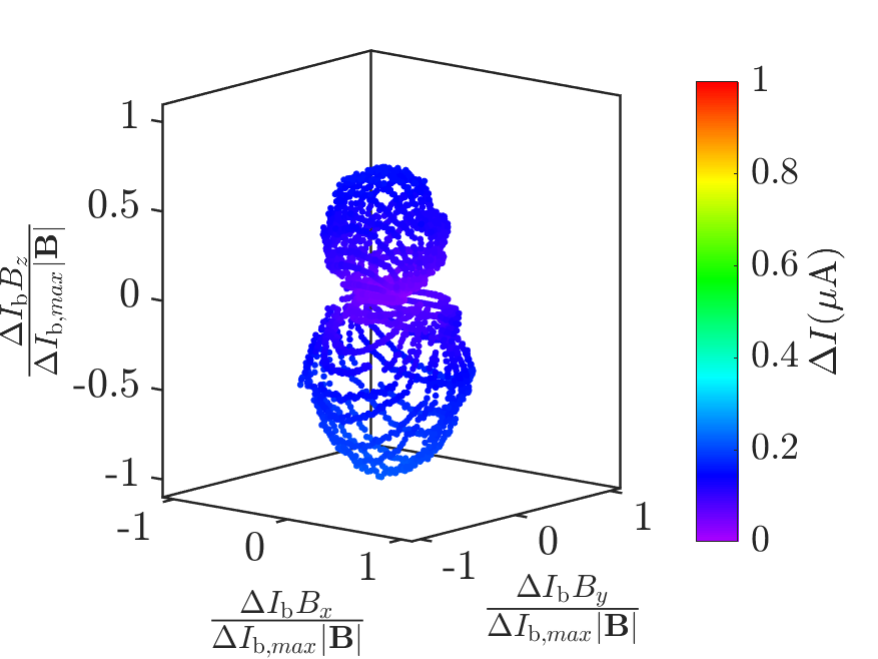}};
    \begin{scope}[x={(image.south east)},y={(image.north west)}]
         \draw[thick, blue] (0.58,0.82) -- (0.58,0.7) -- (0.68, 0.7) -- (0.68, 0.82) -- (0.58, 0.82);
         \node[blue] at (0.63, 0.65) {40$\upmu$m};
    \end{scope}
\end{scope}
\end{tikzpicture}
\caption{\label{fig:dir40p}}
\end{subfigure}
\vskip\baselineskip
\begin{subfigure}{0.49\textwidth}
\begin{tikzpicture}
\begin{scope}
    \node[anchor=south west,inner sep=0] (image) at (0,0) {\includegraphics[width=\textwidth,angle=0]{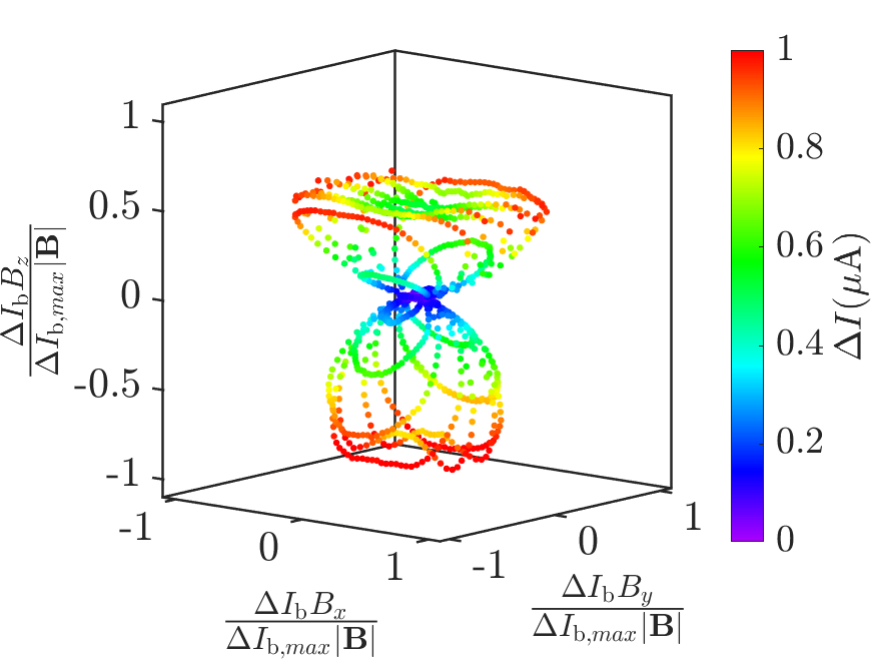}};
    \begin{scope}[x={(image.south east)},y={(image.north west)}]
         \draw[thick, blue] (0.63,0.82) -- (0.63,0.7) -- (0.73, 0.7) -- (0.73, 0.82) -- (0.63, 0.82);
         \draw[thick, blue, fill=blue] (0.645,0.7) -- (0.645,0.79) -- (0.66, 0.79) -- (0.66, 0.7);
         \draw[thick, blue, fill=blue] (0.7,0.7) -- (0.7,0.79) -- (0.715, 0.79) -- (0.715, 0.7);
         \draw[thick, blue, fill=blue] (0.6725,0.82) -- (0.6725,0.73) -- (0.6875, 0.73) -- (0.6875, 0.82);
         \node[blue] at (0.68, 0.65) {10$\upmu$m};
    \end{scope}
\end{scope}
\end{tikzpicture}
\caption{\label{fig:dir103b}}
\end{subfigure}
\hfill
\begin{subfigure}{0.49\textwidth}
\begin{tikzpicture}
\begin{scope}
    \node[anchor=south west,inner sep=0] (image) at (0,0) {\includegraphics[width=\textwidth,angle=0]{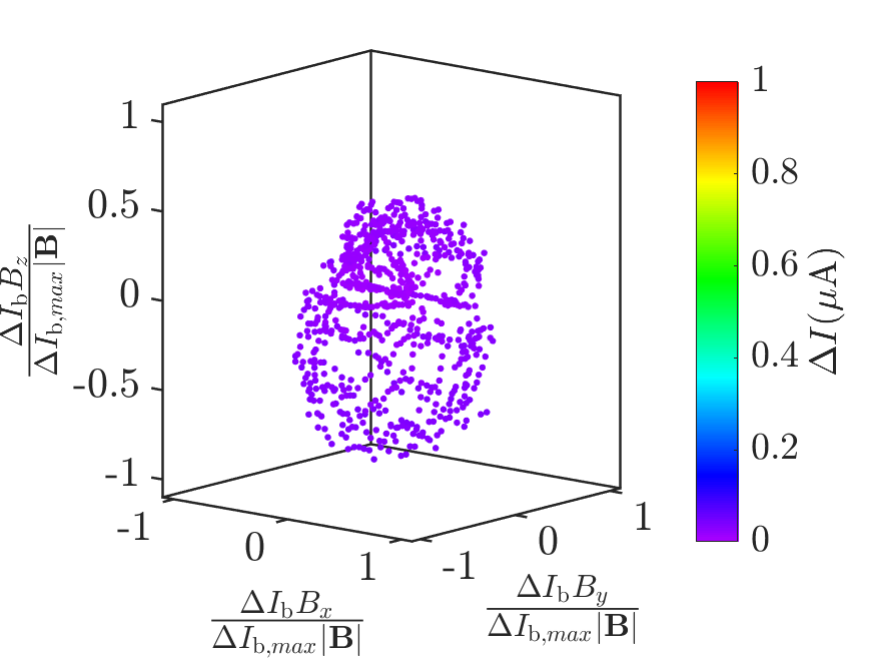}};
    \begin{scope}[x={(image.south east)},y={(image.north west)}]
         \draw[thick, blue] (0.58,0.82) -- (0.58,0.7) -- (0.68, 0.7) -- (0.68, 0.82) -- (0.58, 0.82);
         \draw[thick, blue, fill=blue] (0.595,0.7) -- (0.595,0.79) -- (0.61, 0.79) -- (0.61, 0.7);
         \draw[thick, blue, fill=blue] (0.65,0.7) -- (0.65,0.79) -- (0.665, 0.79) -- (0.665, 0.7);
         \draw[thick, blue, fill=blue] (0.6225,0.82) -- (0.6225,0.73) -- (0.6375, 0.73) -- (0.6375, 0.82);
         \node[blue] at (0.63, 0.65) {40$\upmu$m};
    \end{scope}
\end{scope}
\end{tikzpicture}
\caption{\label{fig:dir403b}}
\end{subfigure}
\caption{\label{fig:dir}
Change in TES bias current with direction of the magnetic field applied.
The change in the measured TES bias current $\Delta I(B)/\Delta I_{\mathrm{max}}$ is plotted as a function of normalised field direction 
for a fixed field $\vert {\mathbf {B}}\vert = 20\,{\rm \upmu T}$.
The blue diagrams in the top right corners of the plots indicate the dimensions and geometry of the bilayer.
This figure shows plots for two unpatterned TESs with (a) a $10\,{\rm\upmu m}$  and (b) a $40\,{\rm\upmu m}$ side length bilayer, and for two TESs with three partial normal metal bars with (c) a $10\,{\rm\upmu m}$ and (d) a $10\,{\rm\upmu m}$ side length bilayer.
These scans were taken at a temperature of 90\,mK and a bias point of 50\,\% $R_{\nn}$. The colours of the points correspond to the magnitude of the current response.}
\end{figure*}
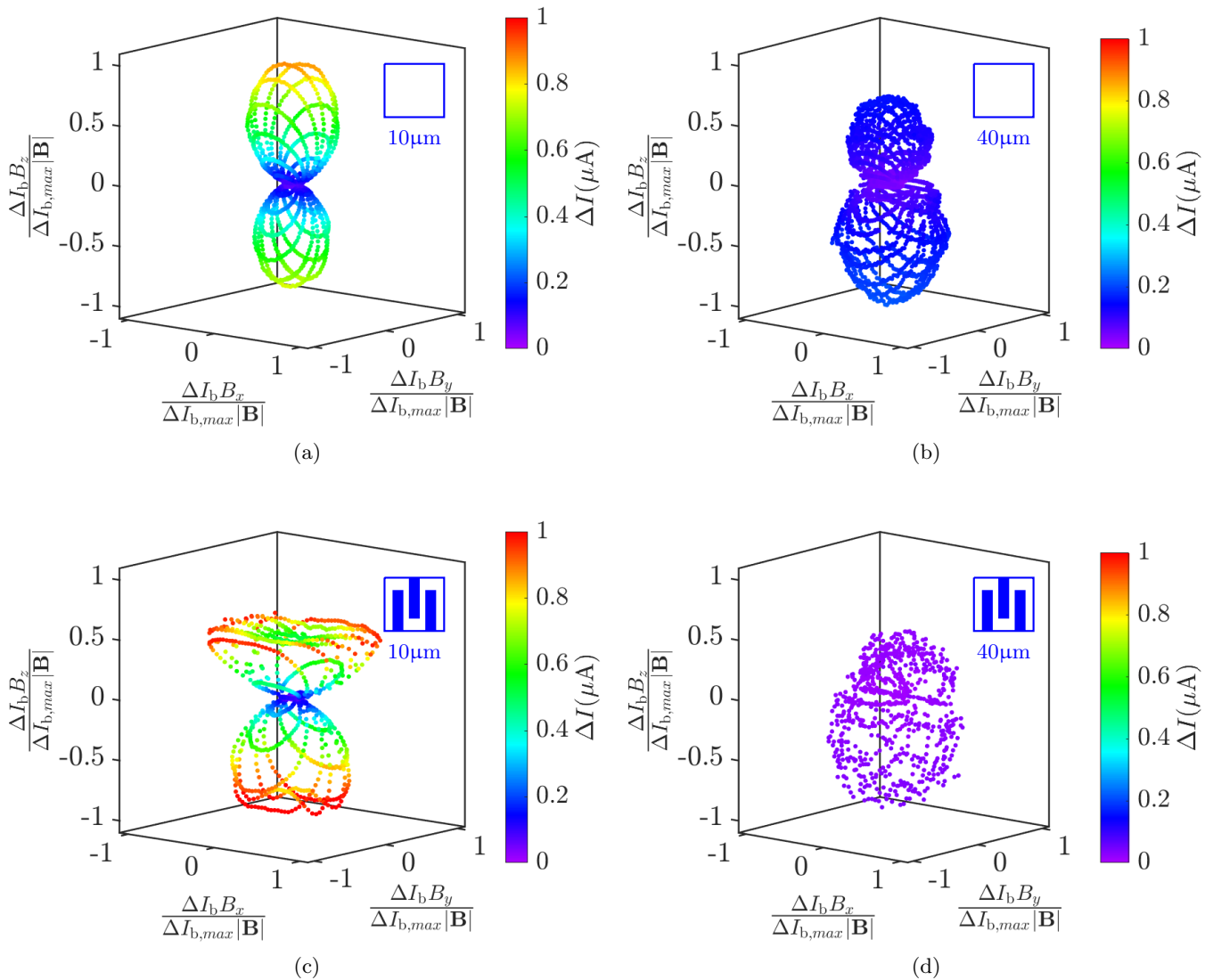

Figure \ref{fig:dir} shows the measured directional dependence of the change in bias current for 
$10\,{\rm\upmu m}$
and 
$40\,{\rm\upmu m}$
TESs either plain or with 3 lateral bars.
The scans were taken at a temperature of 90\,mK and a bias point of  $0.5 R_{\nn}$, with a field magnitude of 
$20\,{\rm\upmu T}$.
In each plot, the change in bias current at a given angle has been normalised by the maximum value recorded for all angles so as to emphasise the directional dependence.
The colour of the points corresponds to the normalised total magnitude of response.
Device geometry is indicated by the cartoon in the top right-hand corner of each plot.
TES size increases across the columns of the figure and the number of bars down the rows.

The smaller unpatterned device shows the simplest directional behaviour: Fig.~\ref{fig:dir10p}.
This has a clean response surface, with a two-lobed form and with no visible oscillations (or ripples) with angle.
The asymmetry in lobe sizes is most likely due to imperfect cancellation of the contribution from the earth's magnetic field and was seen for the other devices.

The larger unpatterned device, Fig.~\ref{fig:dir103b}, displays more complicated behaviour.
Ripples are  visible in the response surface as the field angle is changed.
This links to Fig.~\ref{fig:IB10}, which shows that for changes in magnetic field around and below
$\pm20\,{\rm\upmu T}$
  perpendicular to the plane of the thin film,  $S_I$ (hence $I_b$) changes monotonically with field for an unpatterned TES but that a TES with three normal metal bars is operating close to a local maximum  in $S_I$ (minimum in $I_b$). 
The relatively large oscillations in the response surface for this TES in  an applied field of \SI{20}{\micro\tesla}
 arise from  the local structure in $I_b(B_z)$ as the field direction is changed.

It can be seen by comparing the two rows of Fig.~\ref{fig:dir} that the addition of normal metal bars significantly reduces the directional magnetic field dependence of the TES about the $z$-axis.
This is indicated by the broadening of the main response lobes, which are not described by the simple model in Fig.~\ref{fig:responsesurf}.
In all cases the response in $B_y$ and $B_x$ is below our estimated experimental measurement error.\cite{Harwin2020}

\section{Discussion and Conclusions}\label{Sec:Discussion}

\subsection{Summary of key observations}

The key results  of our measurements  are summarised below, grouped by device parameter. 
For these Mo/Au TESs we observed different behaviour depending on whether the geometric side length was above or below $\approx$20$\,\upmu$m and we distinguish these two types of device as `large' and `small' respectively.
\begin{enumerate}

\item \textbf{Field and temperature dependence of the critical current, $I_c$:}
The behaviour of $I_c$ with field and temperature is important because of its expected influence on TES operating parameters, such as current in bias $I_b$.
All devices showed some degree of Fraunhofer-like fluctuations in $I_c$ with applied  field $B_z$.
The small, unpatterned, devices showed behaviour closest to that of an ideal short Josephson junction demonstrating well-defined zeroes of $I_c$ with field.
Increasing the side-length and adding partial lateral bars were both found to reduce the modulation depth of the pattern.
Very different behaviour was observed in small patterned devices with odd and even numbers of bars that we attribute to the effect of self-field.
In all cases the effective area of the device as determined from the change in $I_c$ with applied flux was smaller than the physical area of the TES.
The temperature dependence of $I_c$ above $T_c$ was well-described by the
description of a SS'S weak-link above $T_{c}$  given in Golubov {\it et al.}\,\cite{Golubov2004} that was derived from the Usadel model. We found a characteristic length  $\xi=450 \pm 20$\,nm that was consistent with the calculated normal-state correlation length of the S' bilayer. The same conclusion was also reached by Sadleir\,\cite{Sadleir2010}.
Below $T_c$, $I_c (T)$  followed the expected Ginzburg-Landau dependence for all TES sizes.

Our measurements of $I_c$ were carried out at fixed bath temperature with $T_{b}\simeq T_{c}$. 
We would expect differences in the magnitude  of  $I_c$  under voltage-biased operation when $T_{b}\sim 0.5T_{c}$, and where Joule heating raises the operating temperature of the TES above  $T_{b}$ such that the actual TES operating temperature may indeed be lower than $T_{c}$.
However, the functional dependence of the behaviour should be similar.

\item \textbf{Field dependence of the operating point of a voltage-biased TES with electrothermal feedback:}
All measurements were made on membrane suspended devices, so as to be representative of sensitive bolometric/calorimetric detectors for low incident powers or energies (and having low noise equivalant powers or high energy resolution respectively).

Superconducting critical temperatures $T_c$ were found to decrease with applied field by about 150\,K/T and 500\,K/T for large and small devices respectively.
For larger TESs, the reduction in $T_c$  could be modelled in terms of the effect of flux-flow resistance using a characteristic length  $\xi = 470 \pm 20$~nm, again consistent with $\xi$ being the normal-state correlation length of the S' bilayer. 

The field dependence of $I_b$ was found to be similar that of $I_c$, consistent with assumptions above.
The Joule power in bias was found to be only weakly dependent on applied field, changing by $\approx 5$\,\% in a 50\,$\upmu$T field for a large TES. 

\item \textbf{Dependence of the operating current, $I_\text{b}$ on field orientation:}
We measured the dependence of the change in $I_b$ on field direction and found significantly reduced sensitivity to in-plane fields, as expected.
We found that smaller TESs were most sensitive to applied fields. 
The addition of lateral bars also reduced the angular dependence. 

\item \textbf{TES power-to-current responsivity with applied field:} 
TES power-to-current  responsivity (change in current for a given change in input power) was shown to vary with field in all devices.
Variations of order 10\% with field were measured for small devices, with little reduction when bars were added.
The variations were reduced in larger devices ($\sim2\%$ for a 40\,$\upmu$m device) and the addition of three bars was found to suppress them nearly entirely.
However, the larger device without lateral bars showed additional, reproducible, discontinuities in sensitivity (and bias current) not present in the small devices, which we attribute to the release or movement of pinned phase slip features. 

\end{enumerate}

\subsection{Recommendations for application of TESs in a magnetic field}
The measurements suggest routes towards optimal TES designs for future applications with respect to the effect of magnetic field.
We will primarily use the behaviour of the bias current $I_b(B)$  hence the power-to-current responsivity  $S_I$ to inform our
recommendations, as it is most directly relevant to detector operation. 
In addition, the effects of the magnetic field on other parameters, like
critical temperature and current, are primarily felt through their
effect on this $I_b(B)$.
In this paper we have not attempted to model the behaviour of $I_b(B)$ in terms of, for example, the resistively shunted Josephson (RSJ) model\cite{Kozorezov2011,Smith2013,Harwin2017} or a two-fluid mode\cite{Bennett_Swetz_2012}
but the key-point is that both require $T_c$ and $I_c$ as inputs in the modelling of $I_b$. 

Based on the observations reported our overall recommendations are listed below: 
\begin{itemize}
\item
Zero field should be the goal in all applications. $B$ and changes in $B$ should be minimized. But this may not be a realistic solution.   
\item
The smallest TESs are least sensitive to changes in field near $B_z=0$. See Eq.~[1]. Our measurements confirm this.
\item
TESs are less sensitive to changes in $B_y$ and  $B_x$ as expected: See Eq.~[1]. Our measurements confirm this. Routing of cabling harnesses or positioning of apertures, for example, should consider this observation.
\end{itemize}
For operation in non-zero  fields with fluctuations (that may occur or be unavoidable in applications) the comments depend on the magnitude of the  fields but generally:
\begin{itemize}
\item
Smaller TESs are more sensitive to changes in applied field, see $S_I(B)$ shown in Fig.~\ref{fig:IBbars}. 
\item
The larger TESs measured here were less susceptible to field (and variations). 
\item
But for larger TESs we also observe discontinuities in $I_b$, hence $S_I$, with $B_z$ that would seriously compromise detector operation within a varying stray field. 
\item
The addition of lateral normal metal bars further reduces the magnetic field sensitivity in $I_b$, hence $S_I$, for larger TESs, and for these TESs no discontinuities in $I_b$  were observed.
\end{itemize}
It is important to emphasise that these considerations do not represent an optimisation for general  TES design. Rather we expect the  issues  to be material, operating temperature and geometry dependent.

\section{Acknowledgments}\label{sec:acknowledgements}

The authors would like to thank the Scottish Microelectronics Centre for growing the SiN/SiO$_2$ membranes for the devices and for providing DRIE.


\bibliographystyle{apsrev4-1}
\bibliography{harwin2023magnetic_ref_abbr}
\onecolumngrid
\twocolumngrid

\end{document}